\documentclass[%
 reprint,
superscriptaddress,
nofootinbib,
 amsmath,amssymb,
 aps,
]{revtex4-2}
\usepackage{graphicx}
\usepackage{dcolumn}
\usepackage{bm}
\usepackage{hyperref}
\usepackage {xcolor}

\begin{document}

\preprint{APS-QS}
\title{Theory and phenomenology of stressed wave-dark-matter soliton} 
 \author{Tzihong Chiueh}%
 \email{chiuehth@phys.ntu.edu.tw}
 \affiliation{%
 Institute of Astrophysics, National Taiwan University, Taipei 10617, Taiwan
 }%

 \affiliation{%
 Center for Theoretical Physics, National Taiwan University, Taipei 10617, Taiwan
 }%
\author{Yi-Hsiung Hsu}
 \email{r07244003@ntu.edu.tw}
 \affiliation{%
 Institute of Astrophysics, National Taiwan University, Taipei 10617, Taiwan
 }%
 
   \affiliation{%
 Center for Theoretical Physics, National Taiwan University, Taipei 10617, Taiwan
 }%
\affiliation{%
 Department of Physics, National Taiwan University, Taipei 10617, Taiwan
 }%
 
\date{\today}

\begin{abstract}
Soliton in the hostile turbulent wave dark matter ($\Psi$DM) halo of a galaxy agitates with various kinds of excitation, and the soliton even breathes heavily under great stress. A theory of collective excitation for a $\Psi$DM soliton is presented. The collective excitation has different degrees of coupling to negative energy modes, where lower-order excitation generally necessitates more negative energy coupling. A constrained variational principle is developed to assess the frequencies and mode structures of small-amplitude perturbations. The predicted frequencies are in good agreement with those found in simulations. Soliton breathing at amplitudes on the verge of breakup is also a highlight of this work.  Even in this extreme nonlinear regime, the wave function perturbation amplitudes are moderate. The simulation data shows a stable oscillation with frequency weakly dependent on the oscillation amplitude, and hints a self-consistent quasi-linear model for the wave function that accounts for modifications in the ground state wave function and the equilibrium density. The mock solution, constructed from the simulation data, can shed lights on the dynamics of the large-amplitude breathing soliton and supports the quasi-linear model, as evidenced by its ability to well predict the nonlinear eigenfrequency shifts and large-amplitude breathing frequency observed in simulations.
\end{abstract}
 \maketitle
 
\section{Introduction}
\label{sec1}
Collective excitations or quasi-particles are common phenomena in many-body physics~\cite{ash,pines}. These emerging phenomena are pertinent to perturbations about the ground state and mediated by interactions with many particles in the system. Interactions with surrounding particles can yield excitations consisting of a mixture of positive and negative energy modes, such as electrons and holes~\cite{efth}. The collective excitation normally creates density perturbation, which produces force perturbation, which then gives rise to velocity perturbation that finally affects density perturbation. The feedback loop proceeds in a self-consistent way governed by coupled equations. A well known example is the longitudinal plasmon in an electrical conductor. Other collective excitations, such as the transverse plasmon, involve current perturbations and the feedback loop can make a bare photon acquire mass in an electron plasma~\cite{pines}. These elemental excitations are often regarded as individual quanta, where different quanta have almost uncorrelated wavefunctions. 
 
On the other hand, the Bose-Einstein Condensate (BEC) system can be described by a classical field, where individual wavefunctions of many particles are coherently summed together, i.e., high occupation numbers~\cite{gri}. Collective excitations in BEC are likewise coherent superpositions of excited eigenstates of individual quanta and may exhibit macroscopic interference patterns~\cite{mew1,mew2}. With attractive interactions, the Bose-Einstein ground state can be non-trivial, i.e., non-uniform~\cite{bra,avd}, and collective excitations are mediated not only by interactions among itself but also with the ground state.    
 
In the astrophysics setting, the simplest version of wave dark matter ($\Psi$DM) is composed of extremely light spin-0 particles of mass $~10^{-22}$ eV~\cite{hu}, and is a BEC system~\cite{sik}. The extremely-light $\Psi$DM in the early universe was in a false ground state (a spatially uniform state) with a very low level of collective excitations.  Gravitational self-interactions can give rise to a negative effective mass to collective excitations immediately before matter dominates the universe, where long wave excitations are unstable to the Jeans instability and short waves excitations are suppressed below the red-shifted Compton length~\cite{zhang1,zhang2,hsu}.
 
As the gravitational instability progresses in the matter dominated era, the long wave perturbations are unstable, much like the particle dark matter, while short waves remain suppressed~\cite{woo,vel1,vel2,moc1,moc2,sch,nor}.  Meanwhile, when the instability enters the highly nonlinear regime, the true ground state can be spatially separated from a cloud of excited states~\cite{schnat}. The phase separated ground state is a soliton located in the galactic core, and the cloud of excitations are identified to be galactic dark matter halo~\cite{schphy}, within which baryons can be gravitationally compressed and build stars.  Since the soliton is physically surrounded by the cloud of random excitations, it is constantly perturbed. When the soliton collective excitation modes resonate with these halo perturbations, these soliton collective modes are greatly excited.  The $l=0$ soliton breathing mode is the most commonly noticeable self-excitation when the soliton resides in a halo, with a sizable ($\sim 50\%$) amplitude at a stable breathing frequency~\cite{osc2,barry,Li}. 
 
The soliton in simulations has also been observed to wander around chaotically kicked by the turbulent halo~\cite{schwal}. Specifically, the interests have been placed on the survivability a compact star cluster supposedly collocated with the invisible soliton of the dwarf galaxy Eridanus II. While a wondering soliton is able to disintegrate the star cluster in a fraction of Hubble time, there are still rooms called for explanation as to why the central star cluster can still survive in the presence of the large-amplitude soliton breathing. If the soliton breathing frequency is in resonance with orbits of the star cluster or a Super-Massive Black Hole (SMBH) inside the soliton, the star cluster may be disrupted and the SMBH kicked away in a short period~\cite{yale}.  Simulations have empirically found that 
the soliton breathes at a stable frequency a few times lower than possible for such resonant interactions to occur, thereby avoiding the catastrophic destruction of star clusters~\cite{barry}. The above results have been derived from empirical or model calculations; they however do not offer useful insights as to why the soliton should breathe at such a low frequency.  
 
A few works have already discussed the soliton breathing frequency in the literature~\cite{osf3,osf2,osf1}.  The only work of relevance addressing the Schroedinger-Poisson equation~\cite{osf1} has provided a quantitative result using the standard quantum mechanics perturbation treatment. This treatment however has a different starting point for the perturbed wave function of the collective excitation and the results are the best very approximate. The present work therefore aims to provide a rigorous theoretical treatment for soliton collective excitation, which is found to be composed of a mixture of positive and negative energy modes, missed in the standard quantum mechanics treatment.
 
Despite this work will primarily focus on the $l=0$ soliton breathing mode, we will also take substantial efforts to address the unusual $l=1$ dipole excitation.  Due to momentum conservation, the dipole mode can only undergo peculiar internal counter displacements.  Since the gravitational potential has no dipole far field, the dipole force is entirely confined within the soliton and becomes a short-range force. On one hand this creates a substantial negative energy component. On the other hand, the oscillating short-range force may have profound but under-explored effects on stellar heating or tidal disruption inside the soliton. This prospect provides a considerable incentive for us to look into the dipole excitation.
 
We will also consider the large-amplitude soliton breathing mode.  In particular, we will propose a model, with which the simulation data can provide us a mock solution that sheds lights on what the data tell us. We will also lay out a framework for the self-consistent quasi-linear theory, which leads to second-order modifications to the equilibrium wave function, including nonlinear shifts in eigen-frequency and changes in the mode profile. The mock solution appears to be consistent with the quasi-linear theory in reproducing the measured nonlinear frequency shift and large-amplitude breathing frequency.
 
Below is a short note of our analytical approach to small-amplitude perturbations. Due to the long-range nature of gravitational interactions, eigen-modes of the soliton collective excitation are governed by two coupled second-order integral-differential equations.  Instead of solving the eigen-mode numerically, we will adopt a more comprehensive variational principle approach in this work. By constructing a Hermitian energy integral, we are able to identify relevant physical effects, allowing us to grasp what are at work for the collective excitation. 
 
The plan of this paper is as follows. In Sec.~\ref{sec2} we formulate the equations for the self-consistent collective excitation. Madelung transformation is used to relate perturbed wavefunctions to perturbed fluid variables. The orthogonality condition and the Hermitian nature for perturbed wavefunctions and fluid variables are shown. In Sec.~\ref{sec3}, variational integrals are derived for evaluation of the eigenfrequency of collective excitation.  Mass and momentum constraints are considered to impose conditions for the trial functions in Sec.~\ref{sec4}. In Sec.~\ref{sec5}, the results of energy integral minimization are presented. We then perform simulations of the soliton subject to small-amplitude perturbations to verify our predictions presented in Sec.~\ref{sec6}. In Sec.~\ref{sec7}, we extend the simulation investigations to the large amplitude soliton breathing mode. The simulation results suggest a quasi-linear simple harmonic oscillator model in the presence of large-amplitude oscillations, a model that incorporates nonlinear modifications to the equilibrium quantities.  We further discuss possible implications of the results and extensions of the framework developed in this work in Sec.~\ref{sec8}. We conclude the work in Sec.~\ref{sec9}. 
 
In this work, we let $\hbar/m_B=1$, where $m_B$ is the boson mass. We also let $4\pi G=4\pi$, and so whenever $4\pi$ appears it is understood that it means $4\pi G$. In some occasions when confusion may occur, we will spell out $G$ explicitly, with an understanding that $G=1$. In this unit, the peak height of the soliton mass density profile, $\rho_s(r=0)$ is set to $1$, the soliton half-height radius, $r_c=0.69$ and the soliton total mass $M_s=3.9$.  The above information is useful to interpret figures in this work.
 
\section{Fluid Variables, Negative Energy Coupling and Hermitian Property for Collective Excitations}
 \label{sec2}
The excited state eigenfunctions satisfy
$$
H_0 \psi_n^l = \Omega_n^l\psi_n^l
$$
where $H_0$ is the Hamiltonian containing the background potential $V_0$, i.e, $H_0=-\nabla^2/2 + V_0$. (In the present case $V_0$ is the soliton gravitational potential.)   We call these $\psi_n^l$ the background excitation to be distinguished from the collecctive excitation.  
 
On the other hand, the collective excitation describes perturbations around the soliton ground state $\Psi_0$, and the perturbed wavefunction is
$$
\delta\Psi=\Psi-\Psi_0,
$$   
which satisfies the perturbed Schroedinger equation
\begin{equation}{\label{eqn1}}
-i{\partial\delta\Psi\over{\partial t}} = H_0\delta\Psi + \delta V\Psi_0 +\delta V\delta\Psi.
\end{equation}
The last terms on the right is a nonlinear term, and we will drop this term for the following small-amplitude perturbation treatment.
 
While the nonlinear ground state takes the form $\Psi_0=f_0(r)\exp(i\Omega_0 t)$ where $f_0(r)$ is taken to be real and positive without loss of generality and $\Omega_0<0$ when the attractive potential $V_0$ vanishes at infinity, the total wave function under perturbation can be expressed as
 
\begin{eqnarray}{\label{eqn2}}
\Psi & = & \Psi_0+\delta\Psi \nonumber \\
& = &e^{i\Omega_0 t}[f_0+ F\cos(\omega t-m\phi)+iG\sin(\omega t-m\phi)]\nonumber\\
& = &e^{i\Omega_0 t}\left[f_0+{G+F\over 2} e^{i(\omega t-m\phi)} + {G-F\over 2} e^{-i(\omega t-m\phi)}\right],\nonumber\\
\end{eqnarray}
where $G(r,\theta)$ and $F(r, \theta)$ are space-dependent single-valued real functions when $\omega$ is real. Here $\omega$ is the density oscillation frequency, and $m\phi$ the phase, where $\phi$ is the azimuthal space angle with an integer $m$.
The important coupling variable $G-F$ is manifestly shown here but absent in standard quantum mechanics perturbation calculations.  This new variable $G-F$ pertains to the negative energy (relative to the ground state) mode amplitude and is present when the self-consistent coupling $\delta V$ is turned on. From Eq.~\ref{eqn2} it is apparent that the collective excitation is different from simply beating of the background excitations.  In addition, since $\delta V$ also oscillates with the same frequency as $\delta\Psi$, treating $\delta V$ as in the traditional first order perturbation theory in quantum mechanics is dubious.
 
It is easy to see that the $\delta V=v(r,\theta)\cos(\omega t -m\phi)$ with a real $v$ is consistent with Eq.~\ref{eqn1}.
Following Eq.~\ref{eqn1}, the perturbed $F$ and $G$ obey
 
\begin{equation}{\label{eqn3}}
\omega F = (H_0-\Omega_0)G,
\end{equation}
\begin{equation}{\label{eqn4}}
\omega G= (H_0-\Omega_0)F+  v f_0. 
\end{equation}
In the present case and what follows, the Laplacian in the Hamiltonian should be understood as $\nabla^2=\nabla_\perp^2 - m^2(r\sin(\theta))^{-2}$ where $\nabla_\perp^2=
r^{-2}(\partial/\partial r)(r^2\partial/\partial r)
-r^{-2}(\partial/\partial\cos(\theta))(\sin(\theta)\partial/\partial\theta)$in the spherical coordinate.
 
For practicality, we may shift the value of the potential at infinity $V_0(r\to\infty)$ in such a way to make $\Omega_0=0$, which we shall call the proper frame of reference.  The background excitation now satisfies
\begin{equation}{\label{eqn5}}
H_0 \psi_n = (\Omega_n-\Omega_0)\psi_n\equiv\omega_n\psi_n.
\end{equation}
Note that the eigenvalue $\omega_n$ is now a measurable physical quantity, the energy level difference $\Omega_n-\Omega_0$.  Meanwhile in the proper reference frame, the collective excitation satisfies 
\begin{equation}{\label{eqn6}}
\omega F = H_0 G
\end{equation}
\begin{equation}{\label{eqn7}}
\omega G = H_0 F + v f_0.
\end{equation}
The frequency $\omega$ is instead unaffected by the frequency shift $\Omega\to 0$ and also a measurable physical quantity describing the oscillation frequency of the perturbed density. In the following analysis, we will adopt the proper reference frame unless otherwise explicitly mentioned.
 
The representation of the wave function $\Psi=\Re\Psi +i \Im\Psi$ can be regarded as the ``Cartesian-coordinate" field representation in analogy to the complex coordinate $z=x+iy$.  A ``polar-coordinate" field representation analogous to $z=r e^{i\theta}$ has the form $\Psi=f \exp(iS)$, where $f\ge 0$ is the amplitude and $S$ the phase.  This representation lays the foundation for the Madelung transformation for the fluid description of wave mechanics.  The fluid density is $\rho=f^2$ and the fluid velocity ${\bf u}=\nabla S$ satisfy the density equation
\begin{equation}{\label{eqn8}}
{\partial\rho\over\partial t}+\nabla\cdot(\rho{\bf u})=0,
\end{equation}
and the velocity equation
\begin{equation}{\label{eqn9}}
{\partial {\bf u}\over\partial t}=-\nabla\left[{{\bf u}^2\over 2} - {\nabla^2\sqrt{\rho}\over 2\sqrt{\rho}} + V\right].
\end{equation}
The fluid description will be crucial in bringing out the physical meaning of the variational integrals to follow.
 
The perturbed wave function in the polar-coordinate field representation can be expressed as
$$
\delta\Psi=f \exp(iS)- f_0\exp(iS_0)=(\delta f + i f_0 \delta S) \exp(iS_0),    
$$
where $\Psi_0\equiv f_0 \exp(iS_0)$ and we may choose the constant phase $S_0=0$. We thus identify the perturbed $x$ field component $F\cos(\omega t-\alpha)$ to be the perturbed field ``radius" $\delta f$, and the $y$ field component $G\sin(\omega t-\alpha)$ the perturbed field displacement normal to the radial direction, $f_0 \delta S$.  
The perturbed density $\delta\rho = 2 f_0\delta f = 2f_0F\cos(\omega t-\alpha)$ and the perturbed velocity $\delta{\bf u}=\nabla\delta S=\nabla(G\sin(\omega t-\alpha)/f_0)$.
 
With the identification of variables and the equation for the ground state wave function $H_0 f_0=0$, it is straightforward to show that Eq.~\ref{eqn6} is equivalent to a disguised version of the perturbed density equation in the fluid description, 
\begin{equation}{\label{eqn10}}
{\partial\delta\rho\over\partial t} = - \nabla\cdot (\rho_0\delta{\bf u}),
\end{equation}
and Eq.~\ref{eqn7} is the linearized version of the perturbed quantum Bernoulli's equation
\begin{equation}{\label{eqn11}}
{\partial\delta S\over\partial t} = - {1\over 2}\delta(\nabla^2 \sqrt{\rho}/\sqrt{\rho})+\delta V.
\end{equation}
We will use Eqs.~\ref{eqn10} and ~\ref{eqn11} to shed lights on the proof of the Hermitian property for collective excitation given by Eqs.~\ref{eqn6} and ~\ref{eqn7}.
 
Equations ~\ref{eqn6} and ~\ref{eqn7} can combine in favor of $F$ or $G$ to form an integral-differential equation, and solving this equation poses an eigenvalue problem for $\omega^2$. However, we will not proceed on this route for reasons given in Sec.~\ref{sec1}.  Instead, we will adopt a variational approach, which requires the system to be Hermitian. The Hermitian property of this quantum system turns out to be subtle. A less thoughtful approach to construction of variational energy integrals can be
incorrect, as demonstrated in Appendix~\ref{secA2} of \cite{barry}.
 
As noted earlier, when a self-consistent potential $v$ is present, it will result in a finite $F-G$. Secondly, the variable $F-G$ is a new degree of freedom (c.f. Eq.~\ref{eqn2}), consisting of a negative energy state.  We are so motivated to consider the cross correlation $\Im[\delta\psi] \Re[\delta\psi]$, instead of the self correlation $\Re[\delta\psi]\Re[\delta\psi]$ or $\Im[\delta\psi] \Im[\delta\psi]$ for the system of equations Eqs.~\ref{eqn6} and ~\ref{eqn7}, for exploring the effect of self-interactions.
We also generalize $G$ and $F$ from real to complex functions for the proof.
 
However $\Im[\delta\psi]=G\sin(\omega t-m\phi)$ and $\Re[\delta\psi]=F\cos(\omega t-m\phi)$ have opposite time parity, thus yielding a zero correlation.  To handle this problem, we
define $Q\cos(\omega t-m\phi)\equiv -\int G\sin(\omega t-m\phi) dt= (G/\omega)\cos(\omega t-m\phi)$ so that $Q(=G/\omega)$ has the correct time parity, $\cos(\omega t-m\phi)$. The variables $Q$ and $F$ will be regarded as the primary variables for the proof.  In the following presentation we will not consider the factor $\cos(\omega t-m\phi)$ for the sake of conciseness since all terms in the following integrals will contain this same factor, and $\cos(\omega t-m\phi)^2$ can be factored out. 
 
Multiply Eq.~\ref{eqn7} by $F'^*$ of the same quantum number $m$ and integrate over the entire volume, we have 
\begin{equation}{\label{eqn12}}
\omega^2\int dx^3 F'^*Q=
\int dx^3 [F'^*H_0 F + F'^*f_0 v].
\end{equation}
Exchange primed and unprimed variables and take a complex conjugate of the product to form another equation.  Subtraction of one equation from the other results in
\begin{eqnarray}{\label{eqn13}}
\int dx^3&&(\omega^2 F'^*Q - \omega'^{*2} F Q'^*) =\nonumber\\
& &\int dx^3 [(F'^*H_0 F-FH_0 F'^*) +(F'^*f_0 v-Ff_0 v'^*)].\nonumber\\
\end{eqnarray}
 
We now want to show $\int dx^3 F'^*Q=\int dx^3 F Q'^*$.
Multiply Eq.~\ref{eqn6} by $Q'^*$ and we find the integral
\begin{eqnarray}{\label{eqn14}}
\int dx^3 Q'^* F &=&\int dx^3Q'^*H_0 Q \nonumber\\
&=&\int dx^3\left[\vphantom{\frac{}{}}-{1\over 2}\nabla_\perp\cdot (Q'^*\nabla_\perp Q)\nonumber\right.\\
& &+{1\over 2}\nabla_\perp Q'^*\cdot\nabla_\perp Q\nonumber\\
& &\left.+\left({m^2\over 2r^2\sin(\theta)^2}+V_0\right) Q'^*Q\right].
\end{eqnarray}
The first term on the right is a total divergence which can be integrated out.  The symmetry of exchanging primed and unprimed variables and taking complex conjugation yields $\int dx^3 Q'^*F=\int dx^3 QF'^*$.  
 
Next, the right-hand side of Eq.~\ref{eqn13} will be shown to vanish.
Replacing $Q$ by $F$ and with the same symmetry as above, the first bracket on the right of Eq.~\ref{eqn13} is seen to vanish.  The second bracket of Eq.~\ref{eqn13}
needs the help from Poisson's equation $\nabla^2 v=4\pi(2Ff_0)$, where $8\pi(v'^*Ff_0)=\nabla_\perp\cdot (v'^*\nabla_\perp v)-\nabla_\perp v'^*\cdot\nabla_\perp v-(m^2/r^2\sin(\theta)^2) v'^*v$ .  Again the total divergence is integrated out, and the symmetry makes the second bracket also vanish.  We thus arrive at
\begin{equation}{\label{eqn15}}
(\omega^2-\omega'^{*2})\int dx^3 QF'^* =0.
\end{equation}
Either the integral $\int dx^3 QF'^*=0$ when $\omega^2-\omega'^{*2}\neq 0$, or the integral is non-zero when $\omega^2-\omega'^{*2}=0$.  The former shows the orthogonality of $G$ and $F'^*$ for $G$ and $F'^*$ having different eigen-values, and the latter shows that $\omega^2$ is real.
This proves Eqs.~\ref{eqn6} and ~\ref{eqn7} to be Hermitian.
 
When we let $\delta{\bf u}(=\nabla(G\sin(\omega-m\phi)/f_0))={\bf w}_\perp\sin(\omega t-m\phi)+{\bf w_\phi}\cos(\omega-m\phi)$, Eq.~\ref{eqn14} can be expressed as
\begin{equation}{\label{eqn16}}
{1\over 2}\left({\omega\over\omega'^*}-{\omega'^*\over \omega}\right)\int dx^3  f_0^2 {\bf w}\cdot {\bf w'}^*=0
\end{equation}
To show Eq.~\ref{eqn16},
we note
the density perturbation is related to the momentum density perturbation (c.f., Eq.~\ref{eqn10}) via
$
2\omega F f_0  = -\nabla_\perp\cdot(\rho_0 {\bf w_\perp})-\rho_0(m w_\phi/ r\sin(\theta)).
$
Given this relation, we now have
\begin{eqnarray}{\label{eqn17}}
QF'^* &=&-\left({1\over{2\omega\omega'^*}}\right)\left({G\over f_0}\right)\left[\nabla_\perp\cdot(f_0^2{\bf w}_\perp'^*)+{{mw_\phi'^* f_0^2}\over {r\sin(\theta)}}\right]\nonumber\\
&=&\left({1\over {2\omega\omega'^*}}\right)\left[f_0^2{\bf w}_\perp'^*\cdot\nabla_\perp\left({G\over f_0}\right)-\nabla_\perp\cdot(Gf_0{\bf w}_\perp'^*)\right.\nonumber\\
& &\left.-f_0^2{{w_\phi'^* (mG/f_0)}\over {r\sin(\theta))}}\right],
\end{eqnarray}
where the total derivative will be integrated out. Recognizing ${\bf w}_\perp=\nabla_\perp(G/f_0)$ and $w_\phi=-mG/(f_0 r\sin(\theta))$, and applying the divergence theorem, we find 
\begin{equation}{\label{eqn18}}
\int dx^3 QF'^* = {1\over 2\omega\omega'^*}\int dx^3 f_0^2 {\bf w}'^* \cdot {\bf w},
\end{equation}
which\footnote{The fluid approach using the velocity perturbation $\bf{w}$ is valid so long as the background density $f_0^2$ has no null; if it does the fluid approach may fail, Eq.~\ref{eqn18} may not hold and the integral $\int dx^3 QF^*$ may no longer be positive definite since there may be a surface term at the background density null.} yields Eq.~\ref{eqn16} from Eq.~\ref{eqn15}. Equation ~\ref{eqn18} can also be obtained by multiplying the complex conjugate of Eq.~\ref{eqn6} by $Q$ and then performing a volume integral, a straightforward calculation that we will skip here.
 
When the primed is removed in Eq.~\ref{eqn18}, we may identify the physical meaning of $|\omega|^2\int dx^3 QF^*$ to be the fluid flow energy, a positive definite quantity, and $\int dx^3 QF^*$ to be $(1/2)\int dx^3 \rho_0 |{\bf \xi}|^2$, where ${\bf \xi}(={\bf w}/\omega)$ is the displacement field of the fluid element.  
 
It can be verified that neither $\int dx^3 FF'^*$ nor $\int dx^3 GG'^*$ are to vanish when the primed differs from the unprimed, due to the presence of the interaction term in Eq.~\ref{eqn8}. As a result, we cannot stress more on the importance of the self-gravitational interaction
that changes the symmetry structure of equations. 
 
\section{Variational Principle}
  \label{sec3}
 
Given Eq.~\ref{eqn18} and identifying the primed with the unprimed, Eq.~\ref{eqn12} gives
\begin{eqnarray}{\label{eqn19}}
{\omega^2\over 2|\omega|^2}\int dx^3 \rho_0|{\bf w}|^2 &=& \int dx^3 [F^* H_0 F + v f_0 F^*]\nonumber\\
&=&\int dx^3\left[{|\nabla_\perp F|^2\over 2} +V_0 |F|^2\right.\nonumber\\
& &\left.-{|\nabla_\perp v|^2\over 8\pi}+{m^2(|F|^2-|v|^2/4\pi)\over 2r^2\sin(\theta)^2}\right].\nonumber\\
\end{eqnarray}
It represents the energy of the hydrodynamic collective excitation. When $\omega^2>0$, the variables $G, F, {\bf w}$ and $v$ are real functions, but when $\omega^2<0$, all variables become complex.
The last equality of Eq.~\ref{eqn19} demonstrates the familiar form of energy, consisting of 
the quantum energy, the background potential energy and the negative self-interaction gravitational energy, respectively.  The left-hand side, on the other hand, is the flow energy ($\omega^2>0$) or the negative flow energy ($\omega^2<0$), originally absent in the soliton ground state. Equation~\ref{eqn18} implies that generation of the flow kinetic energy comes from competitions of positive work done from the energy supply of quantum excitation and negative work from the energy sink of self-consistent interactions; no matter whether the energy supply or the energy sink is to dominate, the flow energy will be generated.
 
Mathematically, for a system to be Hermitian, it means that the eigenvalue $\omega^2$ is real and bounded from below for which a minimum eigenvalue $\omega_{min}^2$ exists; on the other hand, for a system to be variational it demands that this minimum-energy eigenfunction can be approached by appropriate trial functions.  If the negative potential energy overwhelms the positive background energy, the system can be unstable where $\omega^2 < 0$, and the minimum energy state is the most unstable state.  
If we were to remove the negative self-induced potential energy in Eq.~\ref{eqn18}, the solution to the variational principle should recover the background eigenfunction defined by $H_0\psi_n =\omega_n\psi_n$, and the minimum of $\omega^2$ is $\omega_1^2$ of the lowest energy background excited state.  (The ground state is excluded due to the mass conservation to be shown in the next section.) In this situation the perturbed wave function $\delta \Psi$ becomes the standard bound-state eigenmodes of the Schroedinger equation $F({\bf r}) \exp(i\omega t)$ and $F-G=0$.  On the other hand, for collective excitations we additionally have a negative energy component $G-F$ from the negative gravitational self-interaction energy, yielding $\omega_{min}^2 < \omega_1^2$.  Note that though the eigenmode of collective excitation may be expressed as a mixture of background excitations; however, there are exceptions as will be demonstrated for the $l=1$ mode.
 
Equation~\ref{eqn19} can be recast into an operationally convenient form for variational principle.  We multiply all terms in Eq.~\ref{eqn19} by $|\omega|^2$ and use Eq.~\ref{eqn6} to convert $\omega F$ to $H_0 G$.
The variational principle then becomes
\begin{equation}{\label{eqn20}}
\omega^2 = {B\over A} \ge \omega_{min}^2
\end{equation}
where
\begin{equation}{\label{eqn21}}
B= \int dx^3 [(H_0 G^*) (H_0^2 G) + 8\pi(f_0 H_0 G^*)\nabla^{-2} (f_0 H_0 G)]
\end{equation}
and 
\begin{equation}{\label{eqn22}}
A={1\over 2}\int dx^3 f_0^2 |{\bf w}|^2  = \int dx^3 G^* \omega F=\int dx^3 G^*(H_0 G)
\end{equation}
where $\nabla^{-2}$ is the inversion of the Laplacian operator $\nabla^2=\nabla_\perp^2-m^2/(r\sin\theta)^2$.
The inequality of Eq.~\ref{eqn20} becomes an equality when the trial function is a true lowest energy collective excitation; any mixture with other functions will raise the value of $\omega^2$ above $\omega_{min}^2$.
 
In the discussions to follow this section, we will resume the reality of $G$, $F$ and $\omega$. This is because $\omega^2>0$ for the soliton system.
 
\section{Constraints}
  \label{sec4}
 
In an isolated soliton, conservation of mass and momentum must be strictly obeyed.  Below we show how the conservation laws can be reinforced.
There is however a caveat to the mass and momentum constraints.  For a non-isolated soliton embedded in a halo that can potentially exchange mass and momentum, these constraints may no longer hold.  Despite the caveat, the non-violation of mass and momentum conservation should hold on the dynamical time scale, $O(1/\omega)$.  It is inconceivable that the mass can exchange with the halo at this rapid rate.  For momentum conservation to violate, the external force must be locally strong, comparable to the self-binding force of the soliton, for it to take effects on this time scale. The $\psi$DM halo itself does not provide a force of this magnitude, but a massive star cluster or a supermassive black hole colocated with the soliton may offer such a possibility.
 
\subsection{Mass Constraint ($l=0, m=0$)}
 \label{sec4.1}
We impose the constraint $\delta M=\int d^3 x \delta\rho = 0$ as the integration is over the entire volume, representing that perturbations respect the mass conservation.  We thus want 
\begin{equation}{\label{eqn23}}
\int dx^3 f_0 F=0.
\end{equation}
Only the $l=0$ eigenfunction may violate the mass constraint.
For $l > 0$ excitations, the mass
conservation is guaranteed by the angular integral. 
 
The constraint for $F$, Eq.~\ref{eqn23}, is precisely the orthogonality condition for the background eigenfunctions $\psi_n^0$ satisfying Eq.~\ref{eqn5}.
When $F$ is represented by a mixed modes of $\psi_n^0$ (excluding $f_0(=\Psi_0)$), the mass conservation, Eq.~\ref{eqn23}, is guaranteed.  Here the superscript ``$0$" stands for $l=0$ modes.  
We therefore express the trial function $F$ as a linear combination of $\psi_n^0$ with $n\neq 0$.
 
On the other hand, $G$ can contain the ground state mode.
The $H_0$ operator however projects away
the ground state contribution from $H_0G$ in Eqs.~\ref{eqn21} and~\ref{eqn22}.
As a result, the energy integrals $A$ and $B$ contain no $n=0$ ground state contribution. 
 
\subsection{Momentum Constraint ($l=1, m=0$)}
 \label{sec4.2}
This constraint demands the dipole mode must be such that the mass center position of the soliton remains fixed during the sloshing oscillation. 
Using two concentric mass shells to represent the soliton mass distribution, when the two shells oscillate back and forth and the inner shell moves to the right, the outer shell must move to the left to keep a zero mass center movement. 
 
The dipole excitations for $G$ and $F$ assume the functional forms --- $(x, y, z)K(r)$ for some radial funtion $K(r)$. Those associated with $x$ and $y$ are $m=\pm 1$ modes, and with $z$ the $m=0$ mode. Due to the space isotropy, we shall only address the $z$-direction displacement without loss of generality.
 
The mass center position in $z$-direction for a perturbation is
$\int dx^3 z (2f_0F)\cos(\omega t)$. We let $F=zK(r)$ with some radial function $K$. The angle can be first integrated, $2\pi\int_{-1}^1 \cos(\theta)^2 d\cos(\theta)=4\pi/3$, and the volume integral becomes $\cos(\omega t)(8\pi/3)\int r^4 f_0 K dr$. The mass center position must remain at zero for momentum conservation, and hence the radial integral must be zero.  From the construction of mass conservation, where $\psi_n^0$ is orthogonal to $f_0$, we need $K(r) = \psi_n^0/r^2$, or
\begin{equation}{\label{eqn24}}
F=(\psi_n^0/r)\cos(\theta).
\end{equation}
 
On the other hand, the momentum conservation demands
the momentum density in the $z$ direction $\sin(\omega t)\int dx^3 f_0^2\partial(G/f_0)/\partial z=-2\sin(\omega t)\int dx^3 G (df_0(r)/dr)(z/r)$ must be zero, where an integration by part has been used. We let $G=zK(r)$.  The resulting radial integral becomes $\int r^3 K (df_0/dr) dr=-\int f_0 (d(r^3 K)/dr) dr$. We thus require $r^{-2}d(r^3 K)/dr=\psi_n^0$ for the radial integral to vanish.  Hence
\begin{equation}{\label{eqn25}}
G=r^{-2}\left[\int r^2\psi_n^0 dr\right]\cos(\theta).
\end{equation}
 
Eqs.~\ref{eqn24} and~\ref{eqn25} are incompatible.
Since $\psi_n^0(r=0)$ is finite, the former is a singular solution at small $r$ and the latter a regular solution linear in $r$.  Therefore, we choose the latter. This choice results in a regular $F$ (c.f. Eq.~\ref{eqn6} and Sec.~\ref{sec5.2}),
\begin{equation}{\label{eqn26}}
F={H_0^1G\over \omega}={\cos(\theta)\over\omega}\left[{-1\over 2}{d\psi_n^0\over dr} + {V_0\over r^2}\int r^2\psi_n^0 dr\right].
\end{equation}
This $F$ can also produce a zero mass center displacement as shown in Appendix~\ref{secA1}.
 
The reason we are able to construct regular $F/r\cos(\theta)$ beyond the standard procedure like for the singular $F$ (c.f., Eq.~\ref{eqn24}) and meanwhile make it orthogonal to $f_0$ is that the regular $F/r\cos(\theta)$ lies outside the functional space spanned by the complete set of bound states $\psi_n^0$'s, which diminish exponentially at large $r$, as $F/\cos(\theta)$ vanishes at $r\to\infty$ as $r^{-3}$.  In fact $G/\cos(\theta)$ also vanishes at large $r$ as $r^{-2}$.
 
Unlike the $l=0$ perturbation, terms in the series expansion of $G=\cos(\theta)\sum_n a_n^1 r^{-2}\int r^2\psi_n^0 dr$ are not orthogonal to each other; neither are the corresponding terms in regular $F$.  But this non-orthogonality property does not hamper the construction of trial functions as long as different eigenmodes of collective excitation $G_i$ and $F_j$ are orthogonal when $i\neq j$ (c.f., Eq.~\ref{eqn15}).
 
\subsection{Angular Momentum Constraint ($l=1, m=\pm 1$)}
 \label{sec4.3}
We turn our attention to the total angular momentum, which also needs to be conserved, $d\langle L_z\rangle/dt=0$. Here $\langle L_z\rangle$ stands for the volume average of the angular momentum density fluctuation. The total angular momentum
of the excitation is $\int dx^3  f_0^2 {\hat z}\cdot {\bf r}\times \delta{\bf u}=\int dx^3  f_0^2 r\sin\theta w_\phi \cos(\omega t+\phi)$, where $w_\phi \cos(\omega t+\phi)=(1/r\sin(\theta))(\partial/\partial\phi)[G(r,\theta)\sin(\omega t+\phi)/f_0]$, and is the $\phi$ component of $\delta{\bf u}$. Here $G(r,\theta)=K(r)r\sin(\theta)$ with $K(r)$ being a radial wave function. It follows that $w_\phi=K(r)$.
 
Due to the factor $\cos(\omega t+\phi)$ in the volume integral of $\langle L_z\rangle$, the $\phi$ integration for the total angular momentum yields a zero volume integral. It implies that no net angular momentum is to be excited, and no constraint is to be imposed on perturbations from the angular momentum conservation.

\section{ Trial functions and Results}
 \label{sec5}
To determine precise values of the collective excitation frequencies, the soliton solution must be constructed with high accuracy.  The soliton solution we build satisfies the virial condition as in Sec.~\ref{secA2}, to a high accuracy with errors $< 10^{-4}\%$.
 
\subsection{$l=0$ radial mode}
 
As $G = \sum_n a_n^0 \psi_n^0$, we have $H_0^0 G=\sum_n a_n^0\omega_n\psi_n^0$ and $(H_0^0)^2 G=\sum_n a_n^0\omega_n^2\psi_n^0$ for terms of $n\neq 0$ in the variational integrals.  
The gravitational self-interaction in fact has a convenient form for $l=0$ modes.  The enclosed perturbed mass $\delta M(r)$ within a radius $r$ can be derived from Eq.~\ref{eqn10}, $\delta M(r)=4\pi\int r^2 dr (2f_0F)=(4\pi/\omega)\int r^2 dr [(1/r^2)d(r^2f_0^2 w)/dr] = (4\pi r^2/\omega) f_0^2 w$, yielding the perturbed gravitational force $-dv/dr=-\delta M(r)/r^2=(4\pi/\omega)f_0^2 w$.  The negative self-interaction energy density in Eq.~\ref{eqn19} thus becomes 
\begin{equation}{\label{eqn27}}
-(1/8\pi) \left({dv\over dr}\right)^2 = -\left({4\pi f_0^2(r)\over\omega^2}\right)\left({1\over 2}f_0^2(r) w^2\right).
\end{equation}
The quantity $4\pi f_0^2(r)$ is in fact $\omega_J(r)^2$, with $\omega_J(r)$ being the local Jeans frequency, illustrating the relation between the Jeans frequency and self gravitational interactions.  
 
We find $G=\sum_{n=0}^5 a_n^0\psi_n^0(r)$ can model the spherical symmetric breathing mode well.  
We adopt the Markov-Chain Monte Carlo (MCMC) method for optimization. The square-normalized minimum-energy trial function $G$ is found to have the optimal coefficients shown in Table~\ref{table1}.  
 
The monopole $G$ has a contribution of the ground state $\psi_0^0$, despite that this contribution has no effect on the variational evaluation of $\omega$.  This peculiar feature is only limited to the monopole mode due to the fact that $\nabla(G/f_0)=\nabla[(G/f_0)-S_0]={\bf v}$, and $G=f_0(S_0+\int_0^r dr v_r)$, where ${\bf v}$ is the fluid velocity, and the constant phase $S_0$ is an integration constant that can render a non-vanishing $\psi_0^0$ component in $G$ of $l=0$. The coefficient $a_0^0$ of the $\psi_0^0$ component can be determined by projecting $\omega G$ in Eq.~\ref{eqn7} to $\psi_0^0$. It follows that   $a_0^0=\omega^{-1}\int dx^3 vf_0\psi_0^0/\int dx^3 (\psi_0^0)^2$, taking advantage of the relation $\int dx^3 \psi_0^0 H_0 F=0$.  In the integral, the asymptotic perturbed potential $v(r\to\infty)$ must be set to $0$. 
 
\begin{figure}[htbp]
    \includegraphics[width=8.6cm]{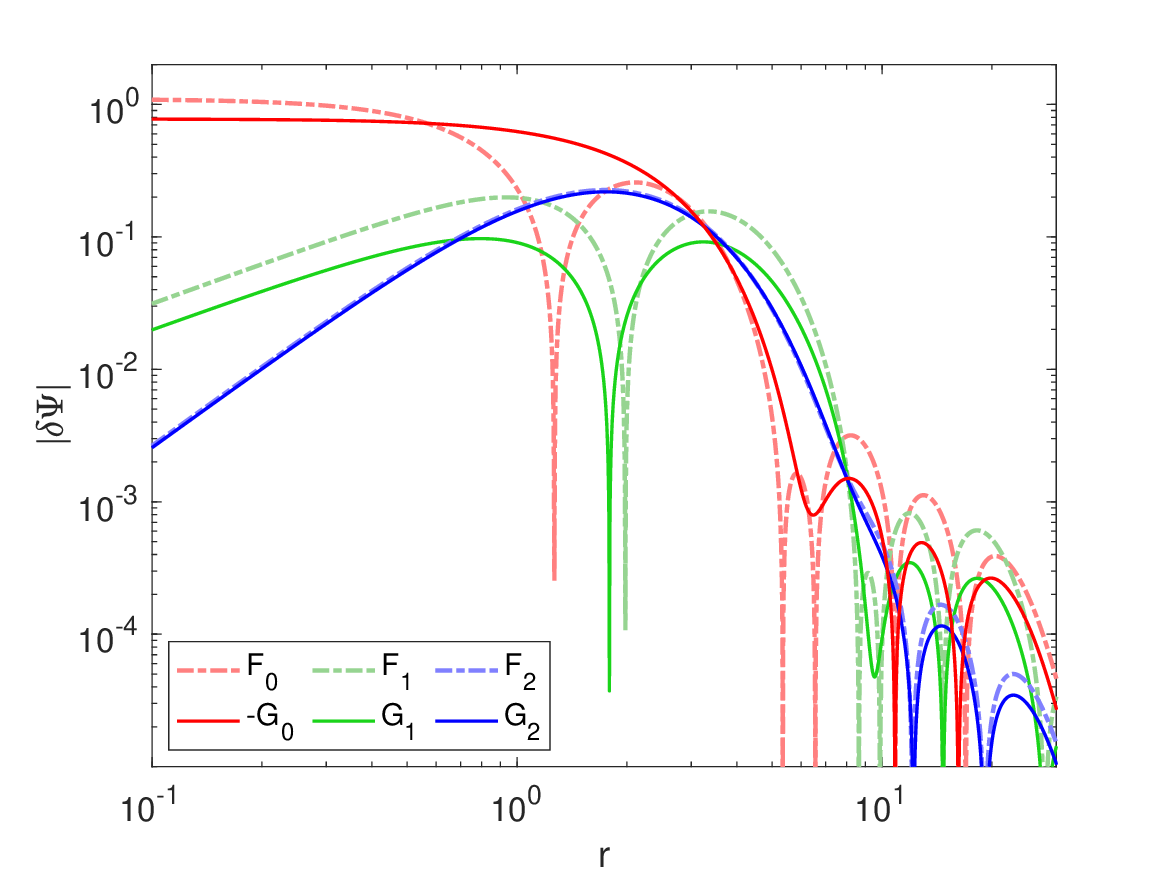}
    \caption{\label{fig:fig1}Radial mode profiles of small-amplitude $F$ and $G$ for $l=0, 1, 2$ collective excitation. (The absolute mode amplitudes are arbitrary.) The monopole mode profiles have the greatest overlaps with the soliton $(r<1)$ and the mode has strong coupling $|G-F/G+F|\sim 600\%$ to the negative energy component.  The dipole profiles have moderate overlaps with the soliton, and it has moderate negative-energy coupling $|G-F/G+F|\sim 30\%$. The quadrapole radial profiles have the least overlaps with the soliton, and it has very small negative-energy coupling as $|G-F/G+F|$ is almost zero.  }
\end{figure}

The trial function $G$ and $F$ are presented in Fig.~\ref{fig:fig1}, where the relative coupling strength of negative-to-positive energy $|F-G/F+G|$ is very large about $600\%$ at the soliton peak.
The frequency so evaluated converges to $\omega_{l=0}= 1.11\sqrt{G\rho_s(r=0)}$, consistent with the breathing frequency $\omega_{l=0}= 1.05\sqrt{G\rho_s(0)}$ found in the soliton simulation to be presented in the next section.  The convergence with respect to the number of terms in the normal-mode expansion is relatively slow, but the agreement is good.
 
Perhaps unexpectedly, the velocity perturbation (=$\nabla(G/f_0)$) in fact diverges exponentially at large $r$.  
The diverging velocity nevertheless yields a
finite flow energy $(1/2)f_0^2 w^2$ as $f_0^2$ decreases with distance much more rapidly to offset the divergence. 
 
\subsection{$l=1$ mode}
 \label{sec5.2}
Due to the constraint of momentum conservation, the $l=1$ modes are the most complicated modes among all to evaluate. Since all $m=0,\pm 1$ modes share the same radial function.  We focus on the $m=0$ mode.
 
We demand $G=Y_0^1 r^{-2}\sum_n\int r^2 \psi_n^0 dr$ for $m=0$, where $Y_0^1=\cos(\theta)$ is the spherical harmonic function of $l=1, m=0$.
This expression of $G$ makes it difficult to evaluate when the $l=1$ Hamiltonian acts on $G$, and it requires some preparatory works.  
 
To begin, we first define $H_0^1$ and $H_0^{1r}$ to be the $l=1$ background Hamiltonian and the radial part of this Hamiltonian, respectively, and $H_0^0$ the $l=0$ Hamiltonian.  For an arbitrary radial function $q(r)$, we have
\begin{eqnarray}{\label{eqn28}}
H_0^1 (\cos(\theta) q(r)) &=&\cos(\theta) H_0^{1r} q(r) \nonumber\\
&\equiv& \cos(\theta)\left[H_0^0 +{1\over r^2}\right]q(r)\nonumber\\
&=&\cos(\theta)\left[\left({-1\over 2r^2}(r^2q(r))'\right)'+V_0(r)q(r)\right].\nonumber\\
\end{eqnarray}
Here, the primed is to denote the radial derivative in this section.
 
Inserting $G$ from Eq.~\ref{eqn25}, we find
\begin{eqnarray}{\label{eqn29}}
\omega F&=&H_0^1G=\cos(\theta)H_0^{1r}\left({1\over r^2}\sum_n \int r^2\psi_n^0 dr\right)\nonumber\\
&=& \cos(\theta)\sum_n \left[{-1\over 2}{d\psi_n^0\over dr} + {V_0\over r^2}\int r^2\psi_n^0 dr\right].
\end{eqnarray}
Using the identity
\begin{equation}{\label{eqn30}}
\int r^2 q(r) H_0^{1r} q(r) dr =\int dr \left[{1\over 2r^2}\left({d(r^2q)\over dr}\right)^2+V_0 r^2 q^2\right],
\end{equation}
we obtain 
\begin{eqnarray}{\label{eqn31}}
\int dx^3GH_0^1G&=&{4\pi\over 3}\sum_n\int_0^\infty \left[ \vphantom{\sum_i}{r^2\over 2}(\psi_n^0)^2\right.\nonumber\\
& &\left.+{V_0\over r^2}\sum_k\left(\int r^2\psi_n^0 dr\right)\left(\int r^2\psi_k^0 dr\right)\right] dr.\nonumber\\
\end{eqnarray}
and 
\begin{widetext}
\begin{eqnarray}{\label{eqn32}}
& &\int dx^3 (H_0^1G) H_0^1 (H_0^1 G) \nonumber\\
&=&{4\pi\over 3}\sum_{n, k}\int dr \left[{1\over 2r^2}\left({r^2\over 2}{d\psi_n^0\over dr}-V_0\int r^2\psi_n^0 dr\right)'\left({r^2\over 2}{d\psi_k^0\over dr}-V_0\int r^2\psi_k^0 dr\right)' \right.\nonumber\\
& &\left.+ V_0r^2\left({1\over 2}{d\psi_n^0\over dr} - {V_0\over r^2}\int r^2\psi_n^0 dr\right)\left({1\over 2}{d\psi_k^0\over dr} - {V_0\over r^2}\int r^2\psi_k^0 dr\right) \right] 
\nonumber\\
&=&{4\pi\over 3}\sum_{n,k}\int dr \left[{1\over 2r^2}\left(\omega_n r^2\psi_n^0 + {dV_0\over dr}\int r^2 \psi_n^0 dr\right)\left(\omega_k r^2\psi_k^0 + {dV_0\over dr}\int r^2 \psi_k^0 dr\right)\right.\nonumber \\
& &\left.+V_0r^2\left({1\over 2}{d\psi_n^0\over dr} - {V_0\over r^2}\int r^2\psi_n^0 dr\right)\left({1\over 2}{d\psi_k^0\over dr} - {V_0\over r^2}\int r^2\psi_k^0 dr\right)\right].
\end{eqnarray}
\end{widetext}
 
The Poisson equation can be treated similarly by recognizing $\nabla^2 (\Re[Y_m^1] q(r))=\Re[Y_m^1](r^{-2}(r^2 q(r))')'$. For the dipole perturbation ($l=1,m=0$), the potential perturbation $v=p(r)\cos(\theta)$, and
we have the Poisson equation
$
(r^{-2}(r^2 p(r))')'=8\pi (f_0/\omega)\sum_n[ (-1/2)(\psi_n^0)' + r^{-2}V_0\int r^2\psi_n^0 dr]$.  Integrate once, and one has
\begin{eqnarray}{\label{eqn33}}
\omega {d\over dr}(r^2 p)&=&4\pi r^2\sum_n \left[\right.-\int  f_0(\psi_n^0)' dr\nonumber\\
& &\left.+2\int {f_0V_0\over r^2}\left(\int r^2\psi_n^0 dr\right)dr + C_n\right],
\end{eqnarray}
where $C_n$ is the integration constant. Integrate one more time and $p(r)$ is identified.
Finally, we arrive at the corresponding term in Eq.~\ref{eqn21} for the self-gravitational energy
\begin{eqnarray}{\label{eqn34}}
& &8\pi\int dx^3 (f_0 H_0^1 G^*)\nabla^{-2} (f_0 H_0^1 G) \nonumber\\
&=&\omega^2\int dx^3 {v\nabla^2 v\over 8\pi} \nonumber\\
&=&-\omega^2\int dx^3 {(\nabla v)^2\over 8\pi} \nonumber\\
&=&-{\omega^2\over 6}\int dr\left[r^2 \left({{dp}\over {dr}}\right)^2+2p^2\right]\nonumber\\
&=&-{\omega^2\over 6}\int dr \left[\left({1\over r}{d(r^2p)\over dr}-2p\right)^2+2p^2\right],
\end{eqnarray}
where Eq.~\ref{eqn33} can be substituted into the last equality, and the left-hand side of this equation is independent of $\omega$ when expressed in $\psi_n^0$. 
 
Note that $p'$ is the perturbed gravitational dipole force and is $\leq O(r^{-3})$.  Hence $(r^2 p)'\leq O(r^{-1})$,
and this fixes the integration constant $\sum_n C_n$ in Eq.~\ref{eqn33}. Moreover, we also need the boundary condition $p\to 0$ for large $r$.
 
Equations~\ref{eqn30}-\ref{eqn34} are ones we have used to calculate the trial function. With $5$-mode series expansion with $\psi_n^0$, the MCMC method finds the optimal trial function $G(= \cos(\theta)r^{-2}\int r^2\sum_{n=1}^5 a_n^1\psi_n^0 dr$) that has square-normalized coefficients given in Table~\ref{table1}.  The optimal trial function gives the eigenfrequency $\omega_{l=1} = 1.718 \sqrt{G\rho_s(0)}$.  As a comparison, the soliton simulation yields an $l=1$ sloshing frequency $1.69\sqrt{G\rho_s(0)}$. The agreement is good. We displays the $l=1$ optimal radial trial functions, $G$ and $F$ in Fig.~\ref{fig:fig1}. The relative coupling strength of negative-to-positive energy $|F-G/F+G|$ is high reaching $33\%$ near the soliton waist.

\subsection{$l=2$ mode}
 
Higher multipole ($l\ge 2$) modes have no constraint from conservation laws.  The trial functions can be evaluated straightforwardly from the linear combination of eigenfunctions of the background Hamiltonian $H_0^l$ as that for the monopole ($l=0$) case, i.e., $G=[\sum_{n=1}^5 a_n^{l=2}\psi_n^{l=2}(r)]\cos(2\theta)$.  The optimal normalized coefficients for $G$ are presented in Table~\ref{table1}, and they are entirely dominated by the first background excitation.  Also shown in Fig.~\ref{fig:fig1} are the optimized $l=2$ radial functions of $G$ and $F$. 
The predicted $l=2$ frequency $\omega_{l=2}= 1.614\sqrt{G\rho_s(0)}$ while the simulation yields $\omega_{l=2}=1.615\sqrt{G\rho_s(0)}$. The convergence of the series expansion is excellent. So is the agreement.  However the relative coupling strength of negative-to-positive energy is essentially zero since $F-G/F+G\approx 0$.
 
\subsection{Low-$l$ Collective Excitation and Trend of High-$l, n$ Modes}
 
Mode structures in Figure~\ref{fig:fig1} reveal that the relevant variable $F-G$ for collective excitation is most prominent for $l=0,1$ modes. The relative coupling strength of negative-to-positive energy $|F-G/F+G|$ can be as high as $600\%$ and $30\%$ for $l=0$ and $l=1$ modes, but the negative energy coupling quickly diminishes for the $l=2$ mode. The negative energy coupling strength is ultimately related to the importance of  self-gravitational interactions, indicating that the $l=0$ mode has the strongest self-gravity effect.
 
On the other hand, from the mode expansion for determining the optimal trial functions, we see the trend for high-$l$ modes.  The MCMC convergence increasingly becomes faster as $l$ gets higher.  The $l=2$ is already almost entirely dominated by the self-gravity-free background state. We therefore anticipate that the background state becomes indistinguishable from even higher-$l$ collective excitation for which a single state dominates.  The trend also applies to those high $n$ modes, for example, the $l=0, n=1$ mode shown to be prominent in Fig.~\ref{fig:fig2} and discussed in the next section.
 
This indicates that high collective excitation modes  become decoupled from the self-gravity. In that situation, the profiles of $G$ and $F$ will be increasingly similar, i.e., $G_l^n(r)\to b_l^n F_l^n(r)$ with $b=\omega/\omega_l^n <1$ but $\to 1$. Here $\omega_l^n$ is the eigenfrequency of the background $n, l$ state. The fact that $b_l^n<1$ is because the frequency of collective excitation $\omega$ is partially contributed by the negative gravitational self-interaction but $\omega_l^n$ has no such contribution.  The reason why $b_l^n\to 1$ for higher $l,n$ can be: higher $n$ means shorter wavelengths, and the self-gravity disfavors short wavelengths, emptying the effect of self-gravity; moreover higher $l$ modes are located too distant away from the soliton, and with a low-density background, self-gravity never prevails.
 
In fact, it does not take high values of $n$ and $l$ to make collective excitation and background excitation indistinguishable. 
For example, the soliton ground state is quantized at about half of the potential depth from the bottom of the potential well; the soliton $l=0, n=1$ excitation is quantized at about $3/5$ higher than the ground state level, but $l=1, n=0$ and $l=2, n=0$ excitations are quantized at $3/4$ and $8/9$ higher than the ground state, where $100\%$ higher corresponds to the continuum. As a result, the rest of higher excitation are like the Rydberg atom approaching the continuum, and the self-gravity has no role to play.
For this reason, only $l=0$ and $l=1$ modes have mode structures of sufficiently long wavelengths and are located well within the soliton to couple to the self-gravity.

\begin{table*}
\begin{ruledtabular}
\caption{\label{table1}Linear Predictions and Simulations}
\label{prediction}
\begin{tabular}{@{}llrc@{}}
\\ \hline\hline
l-mode  & simulation $\omega$\footnotetext[1]{All $\omega$'s are normalized to $\sqrt{G\rho_s(r=0)}$}\footnotemark[1]  & predicted $\omega$\footnotemark[1]  & ($a_0, a_1, a_2, a_3, a_4, a_5)$\footnotetext[2]{All sets of coefficients are squared normalized}\footnotemark[2]
                   \\\hline
0       & 1.05       & 1.11      & (-0.90, 0.424, 0.07, 0.033, 0.021, 0.016 ) \\\hline
1       &  1.69      & 1.718      & (X, 0.841, 0.525, 0.12,0.044,0.012) \\\hline
2       & 1.614      & 1.615    & (X, 0.999, 0.04, 0.02, 0.015, 0.005) \\\hline\hline
                        
\end{tabular}
 \end{ruledtabular}   
\end{table*}

\section{Simulations and Verification of Linear Analyses}
\label{sec6}
We conduct a series of simulations to verify the predicted frequencies of $l=0, l=1$ and $l=2$ modes. 
The simulation code is a part of the GAMER package \cite{gamer1,gamer2} and it runs in a periodic box of $512^3$ grids. Details of our simulation code is largely described in \cite{woo}.  This is a spectral code evolving kinetic energy and potential energy (gravity) separately, where the propagator of the former acts in the Fourier space and that of the latter in the configuration space. As these simulations aim to check against analytical results to a high precision, we choose not to adopt grid refinement so as to avoid any numerical error possibly arising from coarse-fine grid interpolations.  The soliton half-height radius $r_c$ is chosen to be $1/64$ of the box size to permit an ample space for the gravity to work properly.
 
The force balance of the initial unperturbed soliton state has been checked by running a long simulation to ensure the soliton remains static. Subsequently, we excite one particular $l$ perturbation in each simulation run, and the excitation amplitude is sufficiently small $(<10\%\rho_s(0))$ to guarantee the perturbation is in the linear regime. Here $\rho_s(0)$ is the equilibrium soliton density evaluated at the peak $r=0$. All initial density perturbations are constructed via the mass conservation equation, Eq.~\ref{eqn8}, with a displacement field $\xi$, which can be tailored to excite a particular mode.  But in this work we choose a rather arbitrary radial profile of the displacement field, with the constraints that $\xi$ smoothly approaches $0$ as $r\to 0$ and $\infty$ so that $\xi$ peaks near the waist of the soliton. In addition, each simulation is run for 200 oscillation cycles to keep the frequency resolution to $< 1\%$.
 
We have adopted different methods for analysing the oscillations of different $l$ modes. For the monopole ($l=0$) perturbation, we take the instantaneous peak density $\rho_{peak}$ of the soliton as the measure of oscillation.  For the dipole mode ($l=1$), the perturbation is along the $z$-direction and we adopt the volume integral $\int dx^3 r\cos(\theta)\rho({\bf x},t)(r^{-1})$ as the measure of oscillations, where $\cos(\theta)$ is the angle between the radial direction and the $z$-direction. Note that the momentum conservation demands the integral $\int dx^3 r \cos(\theta)\rho({\bf x}, t)$ to be constant. The weighting $r^{-1}$ in our diagnostic integral favors sampling the inner part of the soliton, for the reason that the dipole perturbation changes signs near the soliton waist and this weighting avoids cancellation of opposite signs.  As to the quadrapole mode ($l=2$), one principle axis of the $l=2$ perturbation is also along the $z$-direction, and we take the same volume integral of $l=1$, except for replacing $\cos(\theta)$ by $\cos(2\theta)$, as the measure of oscillations.   
 
\begin{figure}[h]
\includegraphics[width=8.6cm]{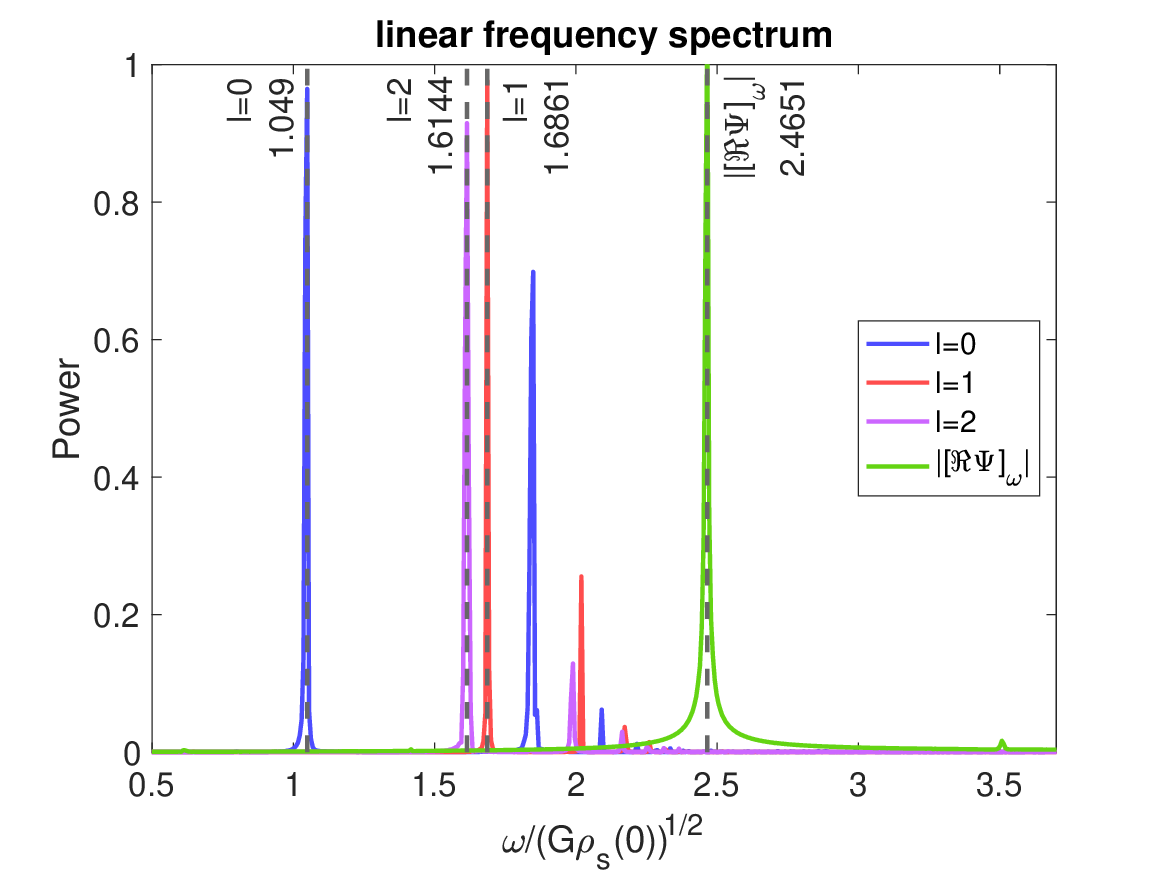}
\caption{\label{fig:fig2}Frequency power spectra
for small-amplitude $l=(0,1,2)$ collective excitation with the primary peaks at around $(1.05, 1.69, 1.61)\sqrt{G\rho_s(r=0)}$, as well as the wave function eigen-frequency response $|[\Re\Psi(0)]_\omega|$ of the $l=0$ mode peaked at around $-2.47\sqrt{G\rho_s(r=0)}$ . (The absolute scale of vertical axis is arbitrary, but the relative strengths of peaks of the same modes are accurate.) All three frequency power spectra of perturbations have distinct secondary peaks near $2\sqrt{G\rho_s(r=0)}$, which are the second or higher excitation of $l=0,1,2$.  The weak secondary peak of $|[\Re\Psi(0)]_\omega|$ at $-3.5\sqrt{G\rho_s(0)}$ comes from the $l=0$ negative energy mode that has frequency $\Omega_0-\omega$, c.f, Eq.~\ref{eqn2}.}
\end{figure}

Spectra of these three sets of time series data are shown in Fig.~\ref{fig:fig2} for $l=0, 1, 2$ modes.  The observed (predicted) $\omega$ normalized to $(G\rho_s(r=0))^{1/2}$ are respectively 1.05 (1.11), 1.69 (1.72), and 1.614 (1.615). The consistent simulation results verify linear perturbation predictions via the variational principle.  
 
The $l=0$ second excitation mode has a sizable peak at $1.83 (G\rho_s(r=0))^{1/2}$. This is in agreement with the predicted value $1.836(G\rho_s(r=0))^{1/2}$ determined by using the variational principle, where
the first excitation mode is projected away from the parameter space of search. Here the convergence of the MCMC search is quite rapid as only $3$ lowest background excitation eigenfunctions are adequate to construct this mode. The second excited modes for $l=1$ and $l=2$ are also visible near $2(G\rho_s(r=0))^{1/2}$ in Fig.~\ref{fig:fig2}. 
 
We also present in Fig.~\ref{fig:fig2} eigen-frequencies of the soliton wave function perturbed by a small $l=0$ mode and evaluated at $r=0$.  Both the real and imaginary parts of the wave function have identical frequency peaks, and we present the frequency response of the real part $|[\Re\Psi(0)]_\omega|$. The main frequency peak at $\Omega_0=-2.465(G\rho_s(0))^{1/2}$ is the negative eigen-frequency of the soliton ground state (c.f. Eq.~\ref{eqn2}).$\footnote{The simulations are not run in the proper reference frame, and $\Omega_0$ is finite and negative}$   A weak secondary peak is near $-3.5 (G\rho_s(0))^{1/2}$, and it corresponds to the negative-energy response of collective excitation $(F-G)/2$ at frequency $\Omega_0-\omega$, according to Eq.~\ref{eqn2}. There should have been another much weaker peak at $\Omega+\omega$ pertinent to the positive energy contribution of $(F+G)/2$, but it is too weak to be visible in Fig.~\ref{fig:fig2}.
 
\section{Large-Amplitude Breathing Mode ($l=0$)}
  \label{sec7}
\subsection{Spectrum Analysis}
 
In most soliton simulations regardless whether the soliton is in isolation or surrounded by a halo, the soliton was seen to stably breathe with a substantial amplitude, which can be as large as $50\%$ about the equilibrium value.  Unexpectedly, the large-amplitude soliton oscillation appears to be non-chaotic with a well-defined frequency.  Motivated by this observation, we set out to conduct a systematic investigation to explore this unique property of the soliton.
 
We begin by conducting a series of simulation runs to examine soliton breathing at different amplitudes.  The simulation setup is identical to the one for small-amplitude $l=0$ perturbation. To control oscillation amplitudes while keeping the soliton mass conserved, we initiate a radial displacement field $\xi(r)$ which yields an initial density distortion $\delta\rho(r)= -(1/r^2)d(r^2\xi(r)\rho_s(r))/dr$.  This ensures
the volume integration of $\delta\rho$ to be zero, thereby conserving the mass.  The choice of $\xi(r)$ is rather arbitrary, except that $\xi(r)$ does not crosses zero, $\xi\propto r$ for small $r$ and $\xi$ vanishes sufficiently fast at a large distance.
 
Three runs are conducted with $30\%$, $75\%$ and $100\%$ density variations ($2(\rho_{max}(r=0)-\rho_{min}(r=0)/(\rho_{max}(r=0)+\rho_{min}(r=0))$). For the case of 100$\%$ density variation, we in fact initialize a displacement field that highly compresses the soliton, yielding an initial density a factor of two higher than the maximum density of the subsequent stable oscillation, but only to find that $10\%$ of the soliton original mass is expelled away to the vast empty space of the simulation box in the first half oscillation cycle and this amount of mass never returns to the soliton. It suggests that 100$\%$ density variation is likely the maximum for the soliton to support stable oscillation.  The fact that very large $l=0$ disturbances leads to mass loss may have an implication of a stable soliton mass for a soliton in a turbulent halo.  We will come back to this issue on the subject of the soliton mass in Sec.~\ref{sec8}.
 
\begin{figure}[htbp] 
    \includegraphics[width=8.6cm]{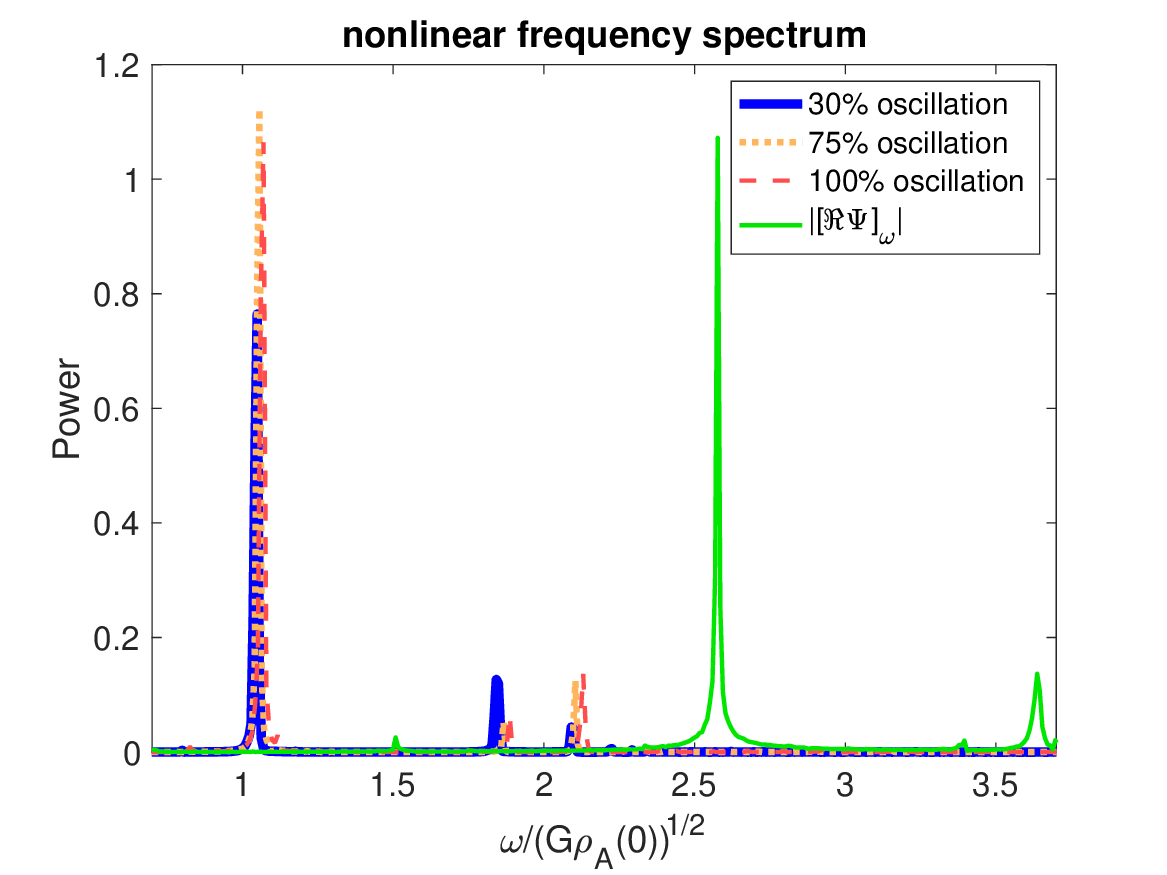}
    \caption{\label{fig:fig3}Frequency power spectra of
$\rho(r=0, t)$ for nonlinear monopole modes of different perturbation strengths and the wave function frequency response $|[\Re\Psi(0)]_\omega|$ for the largest amplitude $l=0$ case. (The scale of the vertical axis is arbitrary.) While the primary frequency peaks of the density oscillations show little changes with the amplitudes, all peaking at around $1.06\sqrt{G\rho_A(0)}$, $|[\Re\Psi(0)]_\omega|$ clearly reveals a frequency shift to $-2.57 \sqrt{G\rho_A(0)}$ compared to the same peak at $-2.465\sqrt{G\rho_A(0)}$ in Fig.~\ref{fig:fig2}. 
The second excitation are heavily suppressed for large amplitude modes; however, the second harmonics of the primary peaks appear at $2.1\sqrt{G\rho_A(0)}$. The secondary peaks of the wave function at $\Omega\pm\omega\approx -2.57\pm 1.07\sqrt{G\rho_A(0)}$ now become clear due to the large oscillation amplitude.}
\end{figure}

The spectrum analysis of three time series of $\rho_(r=0, t)$  surprisingly shows that the soliton breathing modes of different amplitudes all have primary frequency peaks in the narrow range of $1,05-1.07$ $( G\rho_A(r=0))^{1/2}$ as shown in Fig.~\ref{fig:fig3}, where $\rho_A(0)$ is the averaged soliton peak density over the whole run. As the amplitude increases,
the spectral lines are only broadened to a very little extent, indicative of phase stability.
These evidences provide an important clue for our modelling of nonlinear soliton breathing.  
 
The oscillation of $\rho(r=0, t)$ however shows various degrees of amplitude modulation for all cases; the amplitude modulation can be as large as $20\%$ for the largest amplitude case. This amplitude modulation can be caused by the appearance of second harmonics, visible at $2.1( G\rho_A(0))^{1/2}$. The second excitation at $1.83( G\rho_A(0))^{1/2}$ becomes less prominent as the breathing amplitude gets larger.  This observation suggests that we may focus on a single nonlinear mode in modelling the soliton breathing.
 
Finally we show eigen-frequencies of the wavefunction $|[\Re\Psi(0)]_\omega|$ for the $100\%$ amplitude $l=0$ perturbation case in Fig.~\ref{fig:fig3}.  Aside from the dominant negative energy mode pertaining to $G-F$ near frequency $-3.6(G\rho_A(0))^{1/2}$ also seen in Fig.~\ref{fig:fig2}, we now see the positive energy mode associated with $G+F$ near frequency $-1.5(G\rho_A(0))^{1/2}$. 
Upon close examining the main eigen-frequency, we discover a nonlinear eigen-frequency shift $\Delta\Omega$ of the main peak. Comparing Figs.~\ref{fig:fig2} and~\ref{fig:fig3}, the main peak in Fig.~\ref{fig:fig3} shifts to a higher frequency $\Omega_0+\Delta\Omega$ from $\Omega_0$ in Fig.~\ref{fig:fig2}, with
$\Delta\Omega=0.105(G\rho_A(0))^{1/2}$. Since the frequency peak can be identified with high precision, by examining simulation data of lower oscillation amplitudes we find the magnitude of the frequency shift appears to be progressively higher when the oscillation amplitude gets larger. This confirms that the nonlinear frequency shift is real.

\begin{figure*}
\centering
\includegraphics[width=8.6cm]{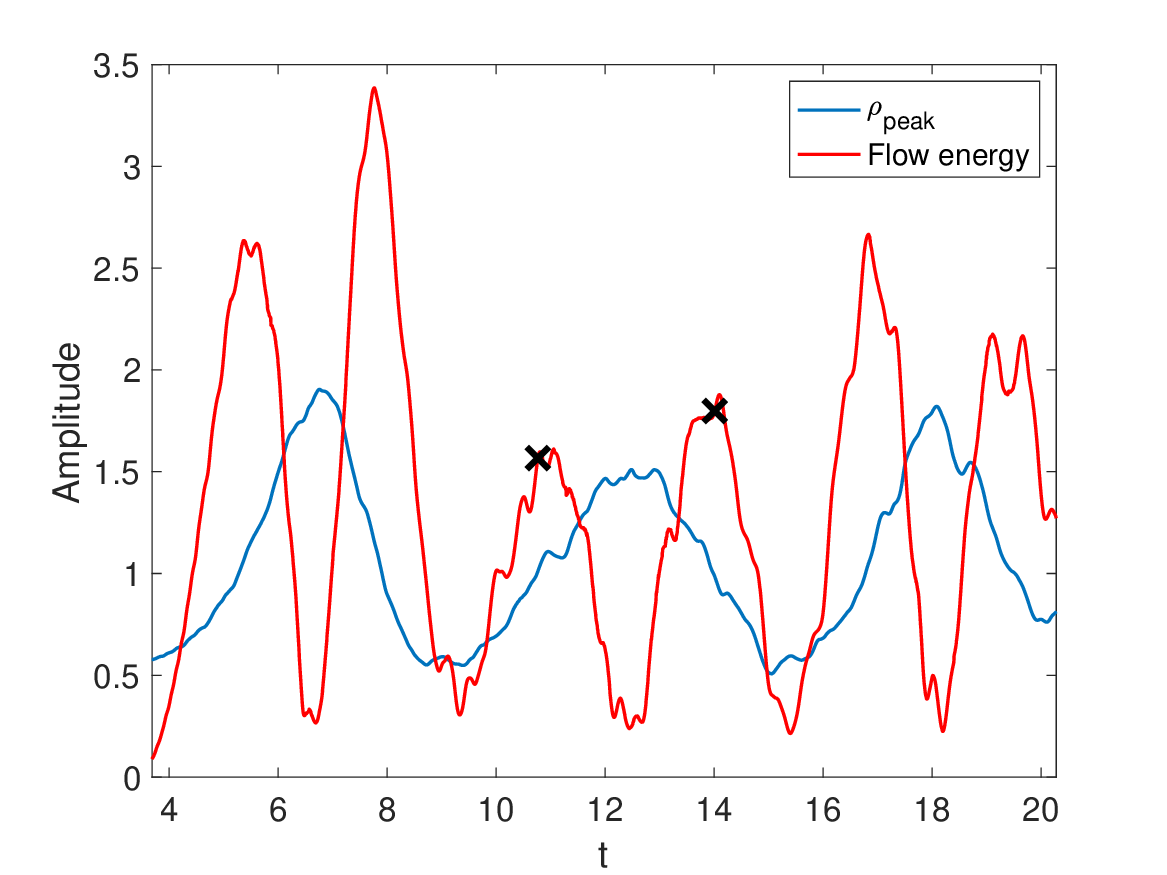}
\quad
\includegraphics[width=8.6cm]{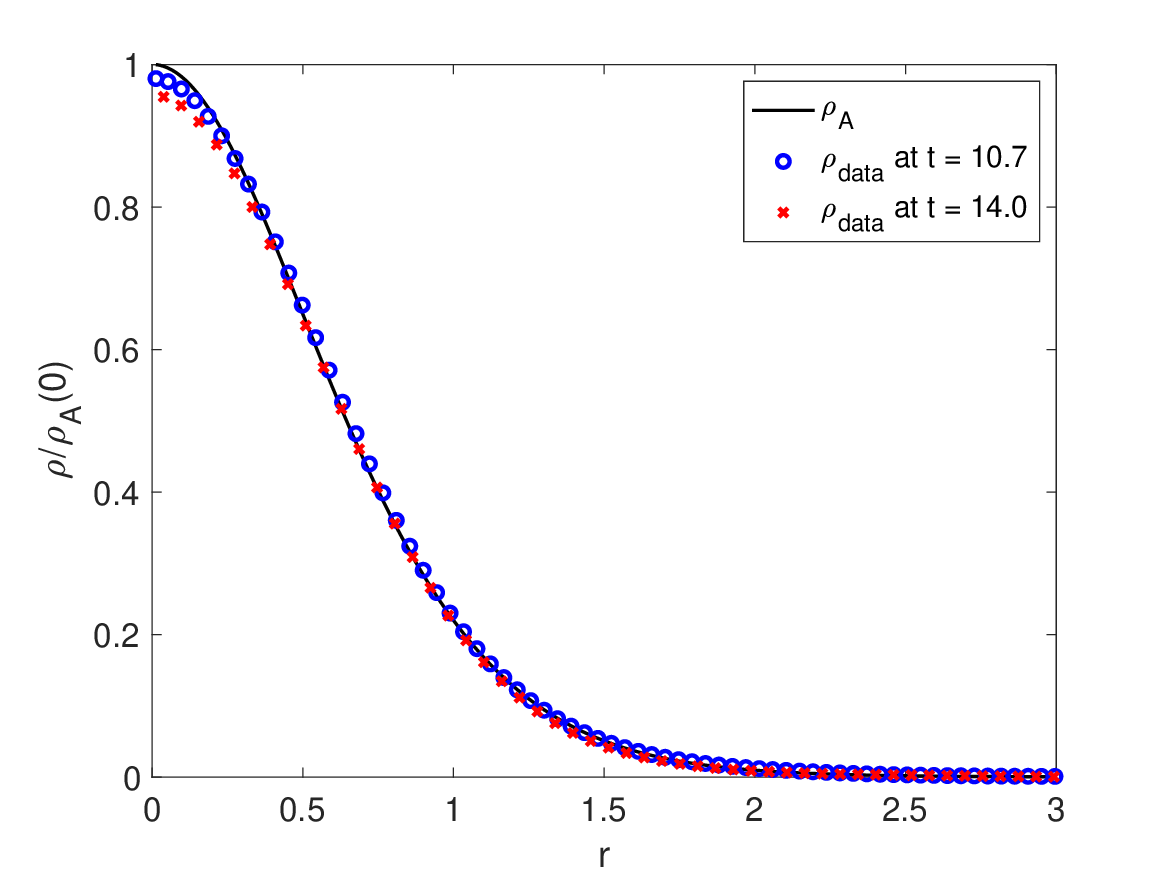}
\caption{\label{fig:fig4}Left: Time series data of the soliton peak density $\rho(r=0, t)$ oscillation and the integrated flow energy oscillation. The density and the flow oscillations are offset roughly by $\pi/2$.  The integrated flow energy appears much more chaotic than the density. This is a result of the inclusion of chaotic flows present outside the soliton near the simulation boundary. The horizonal axis has an arbitrary unit and the vertical scale represents the amplitude of the soliton peak density $\rho(r=0)$. Right: Data of instantaneous density profiles at the two marked points on the left panel of the flow energy maxima.  They are compared with the average soliton density profile $\rho_A(r)(\equiv(\rho_{max}(r)+\rho_{min}(r))/2$), showing remarkable resemblance. This is strongly suggestive of the existence of a stable equilibrium density profile, around which the oscillation proceeds, albeit the nearly $100\%$ large-amplitude oscillation.}
\end{figure*}
 
\subsection{Linear Modelling of Nonlinear Simulation Data}
 \label{sec7.2}
 
We examine the time series data $\rho(t)$ in detail for the most nonlinear (100$\%$ density variation) case.  We showcase the density oscillation for a brief time in Fig.~\ref{fig:fig4}.  Unexpectedly we find that the instantaneous density profile at halfway between the maximum and minimum phases (for example at $t=10.7$ and $14$) of the density remarkably resembles the averaged density profile $\rho_A(r)=(\rho_{max}(r)+\rho_{min}(r))/2$ (c.f., right plot of Fig.~\ref{fig:fig4}).  At this intermediate phase
the oscillating flow energy ($(1/4)\int dx^3 (\Psi^*\nabla\Psi-\Psi\nabla\Psi^*)^2/|\Psi|^2)$ reaches the maximum; on the other hand at the instant the density reaches either the maximum or the minimum, the flow energy is at its minimum.$\footnote{The flow energy oscillation appears to be more random than the density oscillation because the flow energy integral is to a large extent contributed by random flows at great distances.}$
The finding hints
a picture reminiscent of a particle rolling back and forth in a symmetric potential well; the potential minimum at which the velocity reaches maximum is at the equilibrium position.  That is, the averaged density profile $\rho_A(r)$ seems to be the equilibrium density profile, a property that strictly holds for small-amplitude oscillations.  
In addition, we further find that the average profile $\rho_A(r)$ closely resembles the soliton profile $\rho_s(r)$ before the perturbation. Furthermore, a third piece of finding is that the soliton breathing frequency is almost independent of amplitudes.  
 
These
evidences strongly suggest that the large-amplitude oscillation can be treated in a way similar to the small amplitude oscillation, 
and motivate us to propose a quasi-linear simple harmonic oscillator model for large-amplitude soliton breathing. The quasi-linear model proposes that only the equilibrium density and ground state wave function get modified by large-amplitude perturbations, but the dynamics of large-amplitude perturbations remain to be linear. 
 
\begin{figure}
    \includegraphics[width=8.6cm]{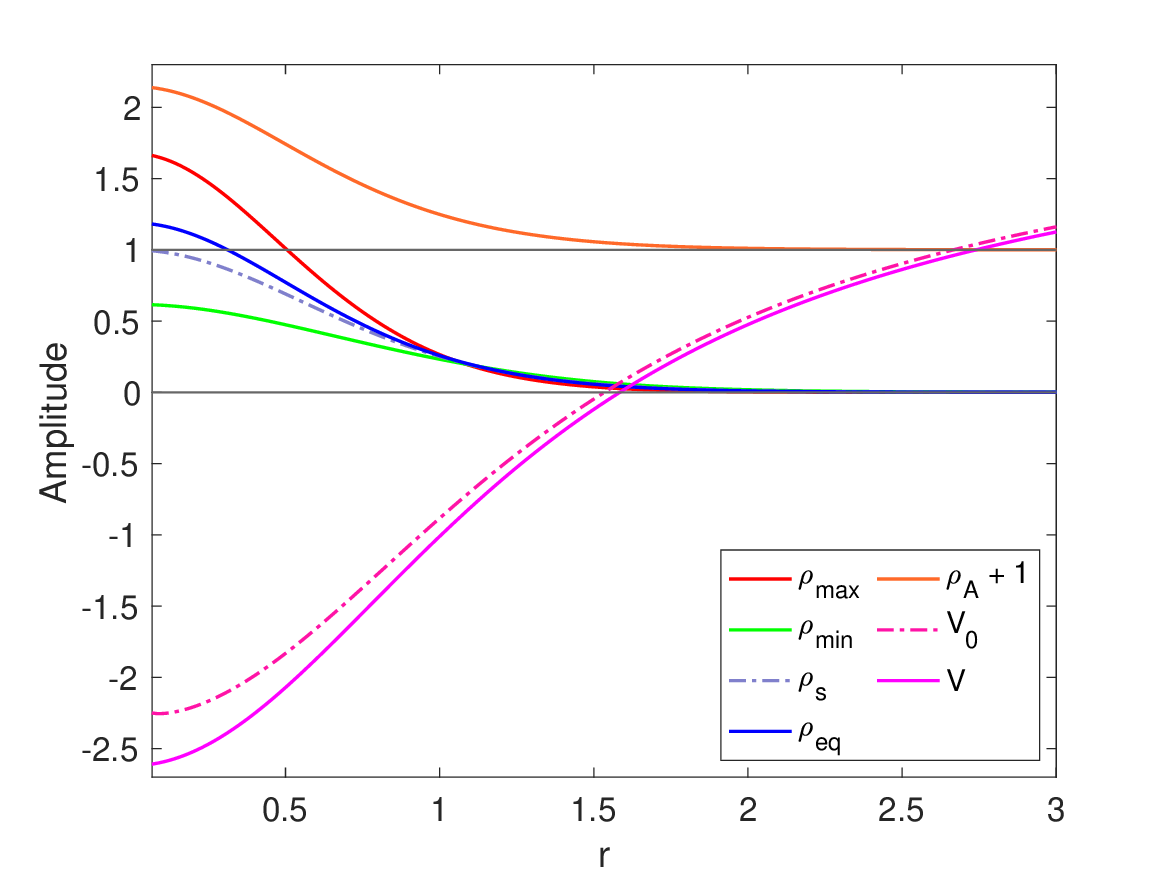}
    \caption{\label{fig:fig5}Data of instantaneous maximum and minimum density profiles, $\rho_{max}(r)$ and $\rho_{min}(r)$, and the unperturbed soliton density profile $\rho_s(r)$.  With the quasi-linear model, the new equilibrium density $\rho_{eq}$ is constructed, whose potential $V$ is to be compared with the unperturbed soliton potential $V_0$. The offset averaged density $\rho_A +1$ is also shown here.  The three densities, $\rho_s(r), \rho_{eq}(r)$ and $\rho_A(r)$, differ from each other by second-order quasi-linear effects of several $\%$ levels.}
\end{figure}
 
However, how can a linear model generate the second harmonic observed in Fig.~\ref{fig:fig3}?
The answer is simple in that the density is a quadratic quantity of the wavefunction, thereby giving rise to the second harmonic, but the perturbed wavefunction remains linear.  
 
Taking into account the frequency shift, the wavefunction is modified to be
\begin{equation}{\label{eqn35}}
\Psi=e^{i\Delta\Omega_0 t}[f_0+ F({\bf r})\cos(\omega t)+iG({\bf r})\sin(\omega t)].
\end{equation}
in the proper reference frame where $\Omega_0=0$. 
The density and momentum density are therefore respectively
\begin{equation}{\label{eqn36}}
\rho=
f_0^2+{1\over 2}( F^2+G^2)+2f_0F\cos(\omega t)+{1\over 2}(F^2-G^2)\cos(2\omega t),
\end{equation}
\begin{eqnarray}{\label{eqn37}}
\rho{\bf v}\cdot \hat r=&&\left(f_0{dG\over dr}-G{df_0\over dr}\right)\sin(\omega t)\nonumber\\
&+&{1\over 2}\left(F{dG\over dr}-G{dF \over dr}\right)\sin(2\omega t).
\end{eqnarray}
Both density and momentum density have second harmonic components that increase with the oscillation amplitudes.
 
Below, we will explore what the simulation data reveals about the new equilibrium wave function $f_0$, finite-amplitude perturbations $F$ and $G$, and new equilibrium density $\rho_{eq}(\equiv f_0^2+(1/2)(F^2+G^2))$ according to the model equations, Eqs.~\ref{eqn36} and ~\ref{eqn37}.
In Fig.~\ref{fig:fig5} we plot $\rho_{max}$ and $\rho_{min}$ next to the marked maximum flow energy phases in Fig.~\ref{fig:fig4}. The two sets of data $\rho_{max}(r)$ and $\rho_{min}(r)$ allows us to solve for $f_0(r)$ and $F(r)$ according to Eq.~\ref{eqn35} at the phases $\omega t=0, \pi$. We also denote $\rho_{\pi/2}(r)$ to be the density at $\omega t=\pm \pi/2$ to solve for $G^2$.
With these data, the solutions are
\begin{eqnarray}{\label{eqn38}}
&&f_0^2(r)={1\over 2}(\rho_A(r) + \rho_G(r)), \nonumber\\
&&G^2=\rho_{\pi/2}(r)- f_0^2(r),\\ 
&&F={\Delta\rho(r)\over [2(\rho_A+\rho_G)]^{1/2}},\nonumber
\end{eqnarray}
where $\rho_A$ and $\rho_G$ are the algebraic and geometric means of
$\rho_{max}$ and $\rho_{min}$, respectively, and $\Delta\rho=(\rho_{max}-\rho_{min})/2$.
 
\begin{figure}
    \includegraphics[width=8.6cm]{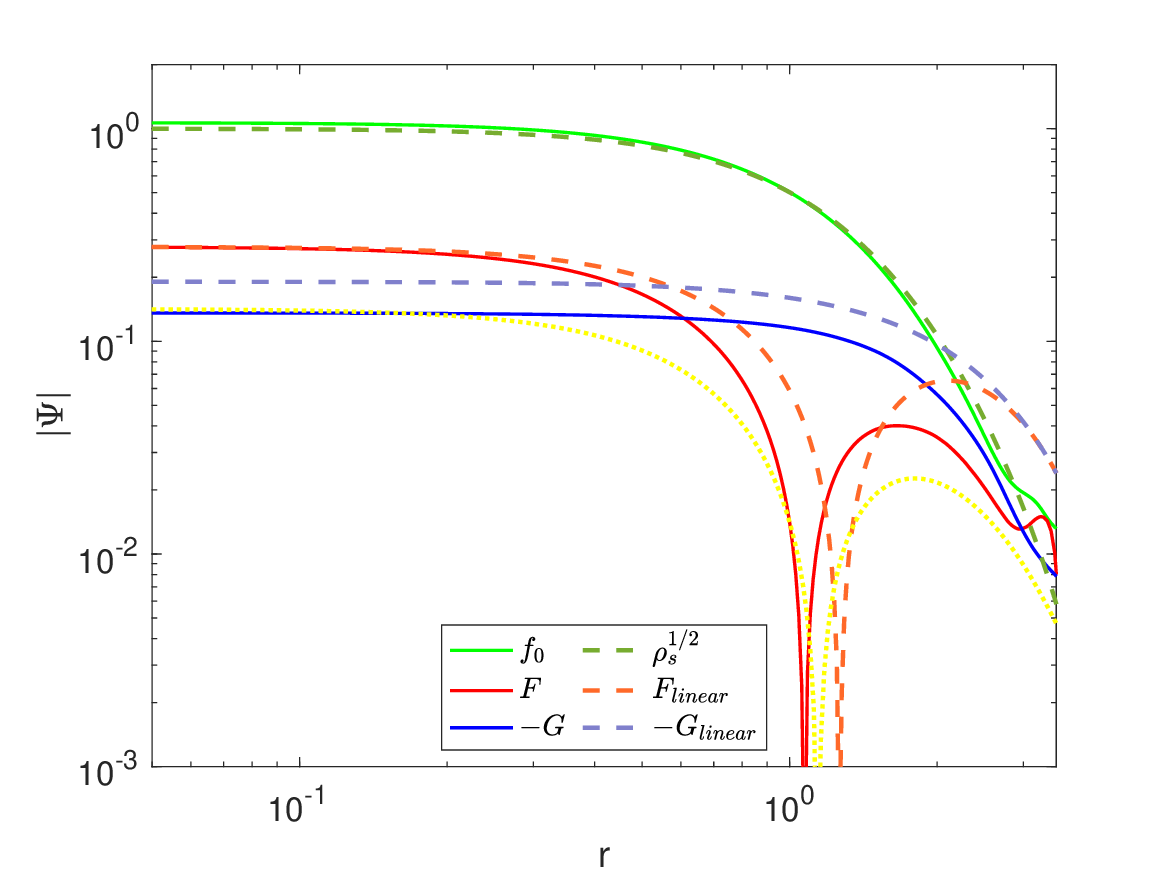}
    \caption{\label{fig:fig6} Nonlinear $F$ and $G$ constructed from simulation data and consistent with the quasi-linear model in comparison with the linear $F$ and $G$.  Also shown are the new ground state wavefunction $f_0$ in comparison with the unperturbed ground state wavefunction $\rho_s^{1/2}$.  The nonlinear amplitude is merely $F/f_0\sim 24\%$, despite $\rho_{max}/\rho_{min}\sim 2.8$. The coupling to the negative energy $|F-G/F+G|$ is comparable to the linear case and can reach $300\%$. As a convergence check, we also construct $F$ of 30$\%$ oscillation amplitude (yellow dotted line), and find that its node lies between $F_{linear}$ and $F$ of 100$\%$ amplitude, indicative of progressively greater modifications of $F$ as amplitudes increase.}
\end{figure}

In Fig.~\ref{fig:fig6}, we plot the perturbation amplitude $F(r)$ in comparison with the new ground state $f_0(r)$.  This is an illuminating figure, as it demonstrates that the amplitude of $F$ is
only $24\%$ of $f_0$, at best a weakly nonlinear amplitude; yet it gives an false impression of a very large density variation $\rho_{max}/\rho_{min}\approx 2.8$. We also plot the linear eigenfunction $F$ for comparison. The nonlinear $F$ has the generic feature of the linear $F$, i.e., sign reversal at about the soliton waist. The $15\%$ narrower nonlinear mode structure suggests a more concentrated density perturbation is capable of driving large-amplitude oscillations at $r=0$.
 
Also plotted in Fig.~\ref{fig:fig6} are the new ground state wavefunction
$f_0(r)$
in comparison with the unperturbed soliton wavefunction $\rho_s^{1/2}$. It reveals that the new ground state wave function $f_0$ is overall smaller than $\rho_s$ at $\%$ level. This is expected because of the mass conservation: $\int dx^3\rho_s=\int dx^3\rho_{eq}$.
 
However there is a problem for $G(r)$ in Eq.~\ref{eqn38} since it can give rise to an unphysical $G^2<0$ at the soliton tail $r>2$.
Unlike data sets $\rho_{max}$ and $\rho_{min}$, which are at stationary phases, the density profile $\rho_{\pi/2}(r)$ not only varies rapidly but also appears unsteady at the soliton tail. This behavior was noticed earlier for the density fits in the right panel of Fig.~\ref{fig:fig4}. 
 
\begin{figure}
    \includegraphics[width=8.6cm]{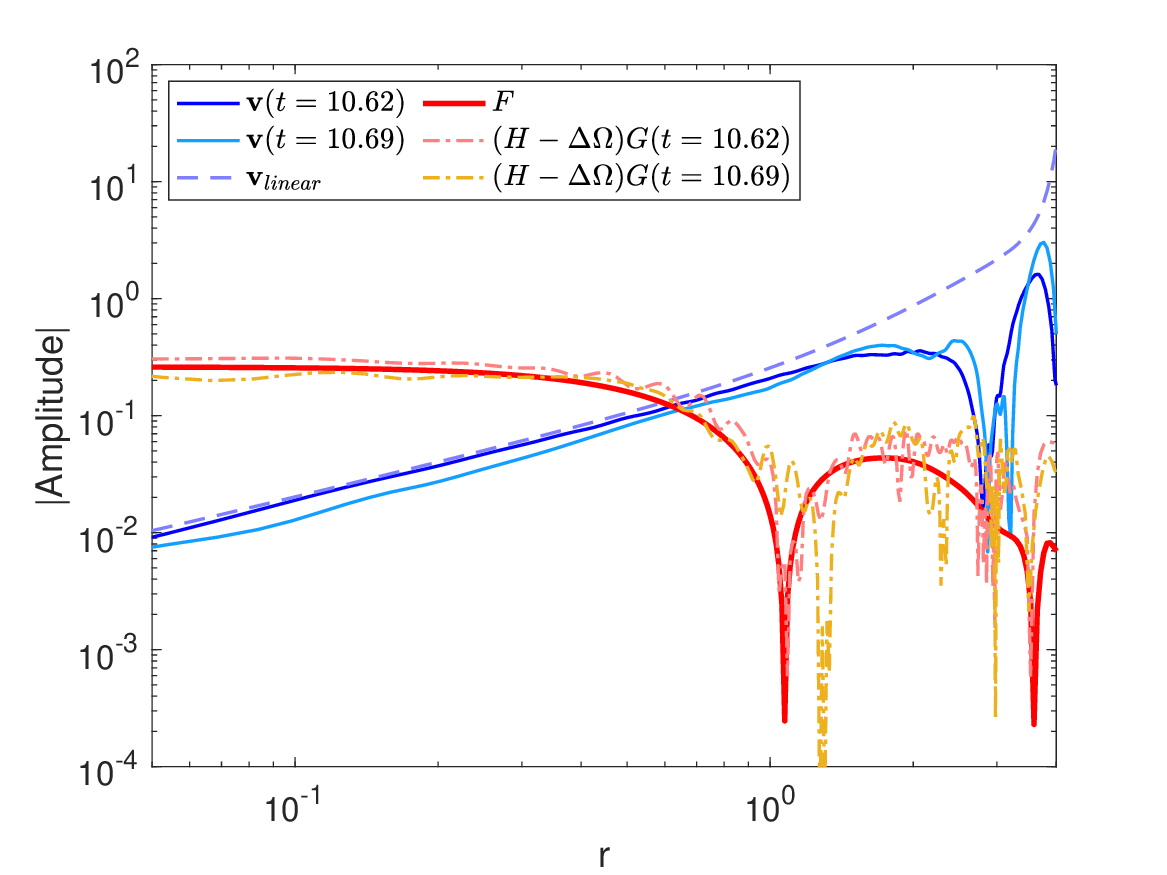}
    \caption{\label{fig:fig7}Nonlinear flow velocity profiles $v(r)$ at their maxima constructed from the simulation data at two adjacent instants (see Fig.~\ref{fig:fig4}). They are to be compared with the flow profile of small-amplitude (linear) perturbation.  From the nonlinear velocity profiles, one can construct two $(H-\Delta\Omega)G$ to be compared with nonlinear $F$ of Fig.~\ref{fig:fig6} .  The ratio of the former to the latter is the nonlinear oscillation frequency $\omega$ according to Eq.~\ref{eqn41}. The best-fit value for the two $\omega$ is $1.08(G\rho_A(0))^{1/2}$, which well agrees with the measured value, $1.07\sqrt{G\rho_A(0)}$, from the simulation (c.f., Fig.~\ref{fig:fig3}).}
\end{figure}

Therefore we seek a different approach to determine $G$.  We adopt the momentum density data at $\omega t=\pi/2$ from the simulation to determine $G$, where 
\begin{equation}{\label{eqn39}}
\rho_{eq} v_r=[f_0^2 d(G/f_0)/dr]\sin(\pi/2),
\end{equation}
according to Eq.~\ref{eqn37}.  The relatively stationary momentum density data at this phase allows us to determine the left hand side reliably.  Performing the radial integration $S(r)\equiv\int_0^r dr (\rho_{eq} v_r/f_0^2)$, we can solve for $G/f_0$. The solution is $G=(S + S_0)f_0$, where $S_0$ is a integration constant. (The velocity data are presented in Fig.~\ref{fig:fig7}, and will be discussed in more detail later in this section.)
 
To fix the magnitude of $S_0$, we resort to $G^2(r=0)$ given in Eq.~\ref{eqn38} with the substitution $\rho_{\pi/2}(r=0)=\rho_A(r=0)$ (c.f., right panel of Fig.~\ref{fig:fig4}). This determines $|S_0|(=|G(r=0))|)$.  The sign of $S_0$ is fixed as follows.
To be consistent with the profile of $G(r)$ in the linear theory,  $S(r)$ (also the radial velocity $v_r(r)$ for this matter) in the main body of the soliton should have an opposite sign to $S_0$ in order for $G(r)$ to possess one node outside the soliton waist.   Inside the soliton, the sign of $G(r)$ is therefore opposite to the sign of $F(r)$, which has the same sign as the radial velocity $v_r(r)$. 
The nonlinear $G(r)$ is shown in Fig.~\ref{fig:fig6}. The relative strength of negative-to-positive energy coupling is large, with $|F-G/F+G|> 300\%$.
 
Having $f_0$, $F$ and $G$ constructed, we now can plot the new equilibrium density $\rho_{eq} (=f_0^2+(1/2)(F^2+G^2))$, c.f., Eq.~\ref{eqn36}, and its potential $V$ in Fig.~\ref{fig:fig5} to show the modifications to the original equilibrium density $\rho_s$ and potential $V_0$.  The new equilibrium density $\rho_{eq}$ has a slightly higher peak but narrower density structure than the original equilibrium density $\rho_s$. In the next two subsections, we will address how to test these mock solutions derived from the simulation data.

\subsection{Self-Consistent Quasi-Linear Modifications of Equilibrium}
 \label{sec7.3}
Only in this subsection will we denote $F$, $G$ and $v$ as $F_\omega$, $G_\omega$ and $v_\omega$ to avoid confusion.
In face of Eqs.~\ref{eqn36} and~\ref{eqn37} for large-amplitude $F_\omega$ and $G_\omega$, the original background wave function $\rho_s^{1/2}$ should be modified to a new $f_0$ and the original Hamiltonian $H_0$ replaced by a new $H$, which satisfy a background wave equation,
\begin{equation}{\label{eqn40}}
(H-\Delta\Omega) f_0=0,
\end{equation}
where $\Delta\Omega$ is a frequency shift, already seen in Sec.~\ref{sec7.2}, due to a change background potential $V=V_0+\Delta V$ in $H$.
Here the background potential $V$ is generated via the Poisson equation with the new equilibrium density $\rho_{eq}$ given in Eq.~\ref{eqn36}. 
The large-amplitude oscillation make the fundamental harmonic mode to satisfy
\begin{equation}{\label{eqn41}}
\omega F_\omega = (H-\Delta\Omega) G_\omega,
\end{equation}
replacing Eq.~\ref{eqn6}.
Likewise, Eq.~\ref{eqn7} becomes
\begin{equation}{\label{eqn42}}
\omega G_\omega = (H-\Delta\Omega)F_\omega + v_\omega f_0.
\end{equation}
 
Contained in Eqs.~\ref{eqn36} and~\ref{eqn37} are also the second-harmonic perturbations $F^{(2)}\cos(2\omega t)$ and $G^{(2)}\sin(2\omega)$. The second harmonic equations 
describing the second-order perturbation can be straightforwardly derived to be
\begin{equation}{\label{eqn43}}
2\omega F^{(2)}_{2\omega} = (H-\Delta\Omega) G^{(2)}_{2\omega}  +{1\over 2} (vG)_{2\omega},
\end{equation}
and
\begin{equation}{\label{eqn44}}
2\omega G^{(2)}_{2\omega} = (H-\Delta\Omega) F^{(2)}_{2\omega} + v^{(2)}_{2\omega} f_0 +
{1\over 2} (vF)_{2\omega}.
\end{equation}
Other than the beating of fundamental modes, the nonlinear terms $vG$ and $vF$ in Eqs.~\ref{eqn43} and~\ref{eqn44} should also include the contributions of high-order harmonic modes, where the beat frequency is $2\omega$.  We show the second harmonic equations not because we intend to address the nonlinear theory, but because Eqs.~\ref{eqn43} and~\ref{eqn44} at zero frequency
can lead to the quasi-linear equation describing how the static background changes as a result of large-amplitude density oscillation.  
 
Taking $\omega=0$, we find Eq.~\ref{eqn43} can produce a static $G^{(2)}_{\omega=0}$ so that the equilibrium wave function changes from real $f_0$ to $f_0 + iG^{(2)}_{\omega=0}$, a phase rotation.  Since the phase rotation is small and difficult to verify from the simulation data, we will not discuss about it.  On the other hand, the $\omega=0$ limit of Eq.~\ref{eqn44} is crucial for the quasi-linear theory, which accounts for the modification of
the ground state wave function from $\rho_s^{1/2}$ to a new $f_0$, and therefore the equilibrium density from $\rho_s$ to $\rho_{eq}(=f_0+(G_\omega^2+F_\omega^2)/2)$ with the new $f_0$. 
 
The zero-$\omega$ limit of Eq.~\ref{eqn44} gives
\begin{equation}{\label{eqn45}}
(H-\Delta \Omega)F_{\omega=0}^{(2)}+v_{\omega=0}^{(2)} f_0  + {1\over 2}v_\omega F_\omega=0,
\end{equation}
where $F_{\omega=0}^{(2)}=f_0-\rho_s^{1/2}$, and $\nabla^2 v_{\omega=0}^{(2)}=8\pi f_0 F_{\omega=0}^{(2)}$.  Keeping all terms in Eq.~\ref{eqn45} of second order, we simplify this equation as
\begin{equation}{\label{eqn46}}
H_0 F_{\omega=0}^{(2)}+v_{\omega=0}^{(2)}\rho_s^{1/2}=-{1\over 2}v_\omega F_\omega,
\end{equation}
where $v_{\omega=0}^{(2)}=8\pi \nabla^{-2}(\rho_s^{1/2} F_{\omega=0}^{(2)})$. This is an integral-differential equation for
$F_{\omega=0}^{(2)}$.  The right-hand side serves as the source to drive the modification $F_{\omega=0}^{(2)}$ that changes
the ground state wave function from $\rho_s^{1/2}$ to $f_0$.
 
How do we know this quasi-linear modification of the ground state wave function is correct?  One self-consistent check is the conservation of the soliton mass:
\begin{eqnarray}{\label{eqn47}}
\int dx^3 \rho_s &=& \int dx^3 \rho_{eq} \nonumber\\
&\equiv& \int dx^3 [(\rho_s^{1/2}+F_{\omega=0}^{(2)})^2 + {1\over 2}(G^2+F^2)],
\end{eqnarray}
accurate up to the second order in oscillation amplitudes.  
 
Now we may count the unknowns and equations.  Given a $\Delta V$, we can solve from Eq.~\ref{eqn40} for the profile of the eigenfunction $f_0$, up to an unknown overall amplitude, and the eigenvalue $\Delta\Omega$.  When $f_0$ is given, we can solve the linear equations, Eqs.~\ref{eqn41} and~\ref{eqn42}, for $F_\omega$ and $G_\omega$, where the amplitude of $f_0$ sets the scale of $\omega$. The solutions $F_\omega$ and $G_\omega$ and $f_0$ in turns will provide an equation for $\Delta V$ since its gravitational source $f_0^2+(1/2)(G_\omega^2+F_\omega^2)-\rho_s$ is now known.  
Lastly, Eqs.~\ref{eqn46} or Eq.~\ref{eqn47} is the equation that closes the loop to fix the unknown amplitude of $f_0$. The whole program involves iteration processes. We will leave the pursuit of this program as a future endeavor.
 
\subsection{Eigen-frequency Shift $\Delta\Omega$ and Soliton nonlinear breathing frequency $\omega$}
 
We will end this section by demonstrating the agreement of the simulation data obained from Sec.~\ref{sec7.2} with the quasi-linear solutions given in Sec.~\ref{sec7.3}. The demonstration focuses on the well-measured nonlinear eigen-frequency shift of the wavefunction $\Delta\Omega$ and the large-amplitude soliton breathing frequency $\omega$.   We shall from now on resume the notation $F$ and $G$ from $F_\omega$ and $G_\omega$.
 
Using Eq.~\ref{eqn40}, where $H=H_0+\Delta V$, treating $\Delta V$ as the potential perturbation, and employing the quantum mechanics first-order perturbation theory, we have
\begin{eqnarray}{\label{eqn48}}
\int dx^3 \rho_s^{1/2}(H_0 + \Delta V) f_0 = \Delta\Omega\int dx^3 \rho_s=\Delta\Omega M_s,
\end{eqnarray}
valid up to second order in oscillation amplitudes,
where $M_s$ is the soliton mass
The first term on the left is zero, and $\Delta V(=V-V_0)$ is given in Fig.~\ref{fig:fig5}. We thus evaluate $\Delta\Omega = -0.11\sqrt{G\rho_{eq}(0)}$ from the quasi-linear theory.  This value is compared with the measured simulation frequency shift obtained from the difference of eigen-frequencies for the $l=0$ modes in Figs.~\ref{fig:fig2} and~\ref{fig:fig3}, and the measured $\Delta\Omega=-0.105\sqrt{G\rho_A(0)}$.  The agreement is very good in view of the tiny shift.
 
As for the $\omega$ determination, we shall explore Eq.~\ref{eqn41}, equivalent to the density evolution equation, to illustrate the mode structures as well as the best-fit value of $\omega$. 
We first check two velocity data from the simulation at the instant near $t=10.7$ in Fig.~\ref{fig:fig4}, where the inward velocities are near their maximum phases, to verify that the nonlinear velocity profiles are largely consistent and also consistent with the linear velocity profile, as shown in Fig.~\ref{fig:fig7}.
We then construct the profile of $(1/2f_0)\nabla\cdot(\rho_{eq}v_r\hat r)=(H-\Delta\Omega)G$, the right-hand side of Eq.~\ref{eqn41}. The profile is also shown in Fig.~\ref{fig:fig7}, which is to be compared with the nonlinear $F$ data, the left-hand side divided by $\omega$. 
The mode structure of nonlinear $(H-\Delta\Omega)G$ are close to that of the nonlinear $F$ in the main body of the soliton, and the ratio of $(H-\Delta\Omega)G$ and $F$ gives the value of $\omega$. 
 
The best-fit $\omega=1.08\sqrt{G\rho_{eq}(0)}$ after averaging the two velocity profiles.  This quasi-linear value of $\omega$ is in good agreement with the measured soliton breathing frequency $\omega=1.07\sqrt{G\rho_A(0)}$ from the simulation. 
 
Alternatively we can evaluate $\omega$ using the energy integral (c.f., Eq.~\ref{eqn19}.  With quasi-linear $f_0$, $\rho_{eq}$ and $H-\Delta\Omega$ in place of $\rho_s^{1/2}$, $\rho_s$ and $H_0$ of the linear theory, we find that the best-fit value of $\omega=1.04\sqrt{G\rho_{eq}(0)}$ by averaging the two $\omega$'s constructed from the two velocity profiles in Fig.~\ref{fig:fig7}.  This value of $\omega$ is also consistent with the measured $\omega=1.07\sqrt{G\rho_A(0)}$ from the simulation.  We conclude that the mock solutions, $F,G, f_0$, constructed from the simulation data are remarkably consistent with the quasilinear theory described by Eqs.~\ref{eqn40}-~\ref{eqn46}.

\section{Discussions and Extensions}
  \label{sec8}
The soliton is a nonlinear ground state of a Bose-Einstein system with negative energy. When sitting in an environment where the surrounding mass is plentifully available, the soliton is expected to grow in mass in time.  However in a halo, the soliton mass is found to be stable, determined by the velocity dispersion or granule sizes of the halo.  
In Sec.~\ref{sec7}, we have found that a soliton can expel mass when subject to a large disturbance.  Therefore the soliton can lose mass when finding itself living in a hostile environment subject to incessant harassment.
Putting together, we come to a suggestion that the soliton should constantly exchange energy and mass with the halo in such a way that it can reach a stable equilibrium, sharing an equal amount of specific energy with the halo (Appendix~\ref{secA2}) and establishing an appropriate amount of its mass.  The mass and energy exchange is through the monopole ($l=0$) deformation and radiation, and this may explain why the soliton in a halo is always observed to breath at a sizable amplitude so as to keep itself in balance.
 
We now in turn address how hostile the soliton can be for stars residing in the soliton. The star orbiting frequency around the soliton is
$\omega_{or}(r)=(4\pi G\langle\rho_s(r)\rangle/3)^{1/2}$, where the angular bracket stands for the volume average. At the soliton center, at which the breathing amplitude is highest, the orbiting frequency is about twice the soliton breathing frequency, $\omega_{or}\approx 2\omega_{l=0}\approx 2(G\rho_s(0))^{1/2}$.
There is no efficient energy transfer mechanism when the donor's frequency is substantially lower than the receiver's frequency.
The particle orbiting frequency can be lower to resonate with the soliton breathing for particle orbits substantially away from the soliton main body.  At the soliton tail, the particle practically only sees a stationary mass at the orbital center in spite of the large-amplitude density oscillation, and no efficient energy transfer is possible either.
 
On the other hand, the soliton $l=1$ internally sloshing mode is different.  The mode has a frequency  $\omega_{l=1}\approx 1.7(G\rho_A(0))^{1/2}$.
The frequency is close to
the orbiting frequency at the soliton's waist, and can approximately resonate with the particle orbit for a range of
radius within the soliton.  Therefore the $l=1$ mode is capable of pumping energy into stars residing in the soliton.  In addition, for a massive orbiting star cluster, the soliton rotating $l=1$ mode can be excited. When
the rotating mode and the star cluster orbit are
co-rotating, some Lagrangian points can be established.  The star cluster may be located near an unstable Lagrangian point, and loosely bound stars outskirt of
the cluster are to be striped away by tidal forces. This is a subject requiring in-depth further studies. What has been said above may also apply to the soliton $l=2$ mode, which has a similar frequency as the $l=1$ mode.  But the $l=2$ mode structure is such that it has too little amplitude interior of the soliton, and hence coupling of the $l=2$ mode and orbiting particles has limited efficiency. 
 
Aside from the specific soliton problem discussed thus far, our main analytical tool, the variational analysis, developed in this paper also works in the context of stability and oscillation of other nonlinear ground states $\Psi_0$.  An obvious example is one with local nonlinearity in the potential $V(|\Psi|^2)$.  The perturbed potential is $\delta V= (\Psi_0^*\delta\Psi+\delta\Psi^*\Psi_0) (dV(f_0^2)/df_0^2) = 2\delta\Psi_r f_0 (dV(f_0^2)/df_0^2)$. The second equality holds after
we adjust the potential level so that the ground state eigenfrequency is zero and $\Psi_0$ is chosen to be real. 
 
Another category of examples is the coupled $\psi$DM with a gravitational potential given by massive baryons or a central massive blackhole.  The time independent ground state $\Psi_0({\bf r})$ may even be allowed to possess a lesser degree of spatial symmetry, such as oval or disk shapes in the presence of an inner stellar bar or disk. Despite the extra complications that may arise, the same Eqs.~\ref{eqn6} and~\ref{eqn7} still holds for the perturbation analysis, where the collective excitation is manifestly Hermitian and the variational principle ensues.  
 
How about the stability of the turbulent $\psi$DM halo itself? In the presence of fluctuating granules and turbulence interior of the halo, the halo is varying on small spatial-temporal scales, but smooth and stationary on large scale.  Therefore the quantum equilibrium resembles an ergodic equilibrium of classical many-particle systems in many ways.   
The statistical description of a turbulent quantum system compared to that of a classical particle system may be understood as follows.  The distribution function $f(E,l)$ of a particle halo can be interpreted as the spectrum of halo granules at energy level $E$ and polar quantum wavenumber $l$ of the $\psi$DM halo\cite{lin}. The equilibrium halo wavefunction is given by $\Psi_{halo}=\sum_{E,l}a(E,l)\psi_{E,l}$, where $\psi_{E,l}$ is the normalized eigen-function, and $a(E,l)$ is the complex random variable of $E$ and $l$ with the squared variance $\langle|a(E,l)|^2
\rangle=f(E,l)$. The stability problem of large-scale perturbations would resemble the problem employing Antosov's variational principle for collisionless particle halos\cite{bin, ant}. We will leave this subject to a future work. 
 
\section {Conclusion}
  \label{sec9}
To conclude, we find frequencies of linear collective excitation determined by the variational principle agree well with the small-amplitude perturbed soliton frequencies from simulations for all three modes, $l=0, 1, 2$, under investigation.
Only the $l=0,1$ collective excitations are found to possess a substantial component contributed by the negative-energy component resulting from attractive self-interactions.  While the mass conservation can be straightforwardly satisfied for all perturbations, the additional constraint of momentum conservation imposed on the dipole mode gives rise to more extended dipole mode structure. This dipole mode is expected to have an appropriate sloshing frequency and spatial distribution to efficiently interact with stars inside the soliton.
 
On the other hand, the simulations of large-amplitude soliton breathing reveal that (1) the maximum breathing amplitude to support a stable soliton is close to $100\%$ density variation relative to the equilibrium density; (2) even though the density variation reaches $100\%$, the perturbed wave function is at best weakly nonlinear with amplitude only $25\%$ of the equilibrium wavefunction, (3)
the frequency of large-amplitude soliton breathing is almost identical to that of small-amplitude oscillation, and only higher by less than $2\%$;
(4) the nonlinearity pertaining to very large-amplitude breathing can modify the ground state wavefunction with an eigenfrequency shift, but both on a level of a few $\%$. 
 
We propose a quasi-linear model to capture the dynamics of nonlinear soliton breathing and derive a system of self-consistent equation.
Despite we have no formal solution to these self-consistent equations, a mock solution constructed from the simulation data and consistent with the quasi-linear model reproduces the measured frequency shift and the breathing frequency to high accuracy.  This provides a strong support to the quasi-linear model for large-amplitude soliton breathing.
 
The nonlinear eigenfrequency shift and the phase rotation of the ground state will never appear in the Madelung fluid variables and equations, and in this regard they are like a gauge degree of freedom. However, in this soliton breathing problem, the fluid variables appear highly nonlinear, but the wave function variables are only weakly nonlinear.  The situation is in great contrast with the cosmological structure formation problem where the background state is uniform.  The wave function variables quickly enter the nonlinear regime beyond the description of the linear perturbation theory, but the fluid variables can remain in agreement with the linear perturbation theory before the density contrast reaches
$>20\%$\cite{woo, li1}. The cosmological significance of this subject lies in the Gaussianity of perturbations\cite{lyn, kun}.  The density perturbation can remain Gaussian random for a long time, but the wave function has already become non-Gaussian early on in the evolution and becomes highly entangled.  The duality nature of fluid and wave function descriptions seems to be related to the relative importance of gravity for the background state. It seems to suggest that strong gravity favors the wave function description, and weak gravity favors the fluid description.  The root of this duality may be profound. 
 
\begin{acknowledgements}
  \label{sec_ack}
We thanks Prof. Hsi-Yu Schive for making GAMER available to us to conduct soliton simulations.  We also thank Barry Chang for his discussions in the early stage of this work.
This project is supported in part by
the grants: MOST-110-2112-M-002-018, and NTU-110-L-890301-3.
 \end{acknowledgements}
 
\appendix
 
\section{Zero Dipole Moment}
  \label{secA1}
We show in Sec.~\ref{sec4.2} that the integrated momentum perturbation must be zero for an isolated soliton.  It requires
the imaginary part of the perturbed wave function to be in a particular form, $G=\cos(\theta)(\int r^2\psi_n^0 dr)r^{-2}$.  To be consistent with Eq.~\ref{eqn7}, the real part becomes $F=\omega^{-1}\cos(\theta)[(-1/2)(d\psi_n^0/ dr) + (V_0/ r^2)\int r^2\psi_n^0 dr]$.  Such an $F$ should also yield a zero mass center displacement, i.e., $\int (2f_0F)zr^2 dr=0$.  We now show that this condition can be satisfied.  Removing the polar angle dependence, the time derivative of the mass center displacement, or the dipole moment,  is
\begin{eqnarray}{\label{a1}}
& &{\omega\over\cos^2(\theta)}\int f_0 F z r^2 dr\nonumber\\
&=&-{1\over 2}\int f_0 r^3 {d\psi_n^0\over dr} dr + \int dr f_0 V_0 r(\int r^2\psi_n^0 dr)\nonumber\\
&=& X + Y,
\end{eqnarray}
where $X$ and $Y$ represents the first and second integrals respectively. From now on we denote primes to be the radial derivatives, and we will employ integrations by part repetitively.  The integral
\begin{eqnarray}{\label{a2}}
X&=&{1\over 2}\int dr (r^3 f_0)'\psi_n^0 \nonumber\\
&=& -{1\over 2}\int dr [(r^3 f_0)'r^{-2}]'\left(\int r^2\psi_n^0 dr\right)\nonumber\\
& =& -\int dr\left[{1\over 2r}(r^2 f_0')'+f_0'\right]\left(\int r^2\psi_n^0 dr\right) 
\end{eqnarray}
and on the other hand
\begin{eqnarray}{\label{a3}}
Y&=&\int dr (f_0 V_0 r)\left(\int r^2\psi_n^0 dr\right)\nonumber\\
&=& \int dr\left[{1\over 2r}(r^2 f_0')'\left(\int r^2\psi_n^0 dr\right)\right],
\end{eqnarray}
as a result of $H_0^0 f_0=0$.
Therefore 
\begin{equation}{\label{a4}}
X+Y=-\int dr f_0'\left(\int r^2\psi_n^0 dr\right)=\int r^2dr f_0\psi_n^0=0,
\end{equation}
the orthogonality condition of $\psi_n^0$ to the ground state.

\section{Virial Theorem}
  \label{secA2}
Here we show the virial theorem for a quantum gravitational bound object.  The conservation of momentum reads,
\begin{equation}{\label{b1}}
{\partial\over\partial t}(\rho{\bf v})+\nabla\cdot\left(\rho {\bf v}{\bf v}+{{\nabla\rho}{\nabla\rho}\over 4\rho}-{1\over 4}\nabla^2\rho{\bf I}\right)= -\rho\nabla V,
\end{equation}
where ${\bf I}$ is the identity tensor.
we next convert the right-hand side $-\rho \nabla V=-(1/4\pi G)[\nabla\cdot(\nabla V\nabla V)+{\bf I}\nabla(\nabla V)^2/2]$ with the Poisson's equation.
Inner product both sides of the equation by ${\bf r}$, a position vector, and then integrate over the whole volume.  We thus arrive at
\begin{eqnarray}{\label{b2}}
{\partial\over\partial t}\int dx^3 \rho{\bf v}&&\cdot{\bf r}\nonumber\\
=-\int dx^3 \nabla\cdot && \left[\rho{\bf v}{\bf v}+{{\nabla\rho}{\nabla\rho}\over 4\rho}-{{\nabla V}{\nabla V}\over 4\pi G}\right.\nonumber\\
&&\left.-\left({1\over 4}\nabla^2\rho-{1\over 8\pi G}(\nabla V)^2\right){\bf I}\right]\cdot{\bf r}.
\end{eqnarray}
Performing integration by part $\int dx^3 \nabla\cdot[...]\cdot{\bf r}=\int dx^3\nabla\cdot([...]\cdot{\bf r})-\int dx^3[...]\cdot(\partial{\bf r}/\partial {\bf r})$, we then remove the total divergence via the divergence theorem. Note that the quantity $\partial{\bf r}/\partial {\bf r}={\bf I}$.  Hence we have 
\begin{eqnarray}{\label{b3}}
{\partial\over\partial t}\int dx^3 \rho{\bf v}\cdot{\bf r}=-\int dx^3 &&\left[\rho{\bf v}\cdot{\bf v}+{{\nabla\rho}\cdot{\nabla\rho}\over 4\rho}-{{\nabla V}\cdot{\nabla V}\over 4\pi G}\right.\nonumber\\
&&\left.-\left({1\over 4}\nabla^2\rho-{1\over 8\pi G}(\nabla V)^2\right)\right].\nonumber\\
\end{eqnarray}
The $\nabla^2\rho$ term can again be integrated out.  The virial theorem follows when the left hand side is zero:
$$
\int dx^3 \left[(\rho{\bf v}^2+(\nabla f)^2)-{({\nabla V})^2\over 8\pi G}\right]=0,
$$
or
\begin{eqnarray}{\label{b4}}
\int dx^3 \left[(\rho{\bf v}^2+(\nabla f)^2)+{\rho V\over 2}\right]=0,
\end{eqnarray}
i.e.,
$$
2KE+PE=0
$$
where $f^2=\rho$.
 
For the soliton, we have no flow velocity ${\bf v}=0$, and $(\nabla f)^2/2$ alone is solely responsible for the kinetic energy.  On the other hand, for a turbulent halo, the flow energy $\rho{\bf v}^2/2$ and the quantum energy both contribute to the kinetic energy.  If the flow energy and the quantum energy are in equi-partition, each component contributes half of the kinetic energy.  If furthermore the solition and the halo are in ``thermal" equilibrium, where the specific energies (total kinetic energy density/density) of the soliton and the halo share the same value, the soliton specific quantum energy will be twice as large as the halo specific quantum energy.  This means that the soliton size ought to be $1/\sqrt{2}$ of the granule size.   ``Thermal" equilibrium with the soliton should holds for inner halo that surrounds the soliton.  So is the mass of the soliton in balance with its surrounding halo, a subject addressed in Sec.~\ref{sec8}.
 
Moreover, the fluid turbulent velocity is given by the gradient of the phase $\delta S$ and the quantum velocity dispersion from the gradient of the amplitude $\delta\ln f$.  If the two scalar $S$ and $\ln f$ are uncorrelated in a turbulent $\psi$DM halo, such a halo contains $2$ ``thermal" degrees of freedom, in contrast to $3$ degrees of freedom in classical collisionless particles.  Hence a ``thermalized" halo has a stiffer equation of state than a ``thermailized" collisionless particle halo and the most likely value of the adiabatic index $\Gamma$ is expected to be $2$.  
 
\bibliography{reference.bib}
 
\end{document}